\documentclass[twocolumn,showpacs,preprintnumbers,amsmath,amssymb,amsmath,amssymb,prb,floatfix,lengthcheck]{revtex4}

\usepackage{graphicx}
\usepackage{bm}
\usepackage{hyperref}
\usepackage{color}

\begin{document}

\title{Coherent properties of nano-electromechanical systems}

\author{G. Piovano$^{1}$, F. Cavaliere$^{1}$, E. Paladino$^{2}$, and M. Sassetti$^{1}$}
 \affiliation{$^1$ Dipartimento di Fisica $\&$ CNR-SPIN, Universit\`a di Genova,Via Dodecaneso 33,
  16146, Genova, Italy.\\
\noindent $^2$ Dipartimento di Fisica e Astronomia, Universit\`a di
 Catania $\&$ CNR IMM MATIS Catania, C/O 
Viale A. Doria 6, Ed. 10, 95125 Catania, Italy.}
\date{\today}

\begin{abstract}
We study the properties of a nano-electromechanical system in the
coherent regime, where the electronic and vibrational time scales are
of the same order. Employing a master equation approach, we obtain the
stationary reduced density matrix retaining the coherences between
vibrational states. Depending on the system parameters, two regimes
are identified, characterized by either ($i$) an {\em effective} thermal
state with a temperature {\em lower} than that of the environment
or ($ii$) strong coherent effects. A marked cooling of the
vibrational degree of freedom is observed with a suppression of the
vibron Fano factor down to sub-Poissonian values and a reduction of
the position and momentum quadratures.
\end{abstract}

\maketitle

\section{Introduction}
Nano-electromechanical systems~\cite{roukes} (NEMS) represent an
intriguing class of devices composed of a nano-mechanical resonator
coupled to an electronic nanodevice. Several examples of NEMS have
been realized, ranging from single oscillating molecules,~\cite{park}
to suspended carbon nanotubes~\cite{sap,let,leroy,lass} and suspended
nano-cantilevers or nano-beams.~\cite{knobel,nanobeams}

\noindent In all these devices, the coupling between mechanical and
electronic degrees of freedom gives rise to peculiar transport
phenomena like Franck-Condon blockade,~\cite{let,flens,koch,mitra}
negative differential conductance~\cite{shen,sap,zazunov,cava,braggio,cava1}
and remarkable noise characteristics.~\cite{haupt,koch} Along with
their outstanding electronic properties, also mechanical ones are of
extreme interest.  Indeed, owing to the extreme sensitivity of the
vibrating part to the spatial motion, NEMS have been proposed as novel
detectors in scanning microscopes~\cite{sid} or as ultra-sensitive
nano-scales, being able to detect the mass of even few molecules
adhering to them.~\cite{il,nanoscale} In order to successfully employ
a NEMS as a precision position detector it is important to reduce its
thermal fluctuations, eventually attaining the ultimate goal of
cooling it down to its quantum ground state.~\cite{giazotto} Also,
ultra-sensitive NEMS position detectors based on peculiar quantum
states such as position-squeezed states~\cite{mandel} have been
proposed and experimentally realized.~\cite{hertz}

\noindent A great variety of NEMS setups have been investigated
theoretically. Typically, the electronic part is composed of a
semiconducting,~\cite{koch2,mozy,nocera,huss,merlo,cava3}
normal~\cite{armour,armour2,doiron} or
superconducting~\cite{rod2,harvey,rodr2,clerk} single quantum
dot.~\cite{kou} Double-dot setups have been studied as
well.~\cite{ouy,huss} Different models for the coupling between
electrons and vibrons have been considered, ranging from the simple
Anderson-Holstein (AH)
model~\cite{koch,koch2,pisto,pisto_c,rodr,nocera,huss,merlo,cava3,schultz}
to microscopic models tailored for specific systems, such as suspended
carbon nanotubes.~\cite{izu,flens2,cava2} Also
the influence of external dissipative baths~\cite{galp3,galp4,cava3}
or radiation fields~\cite{ouy,rae,teufel} has been analyzed. In the
most complex configurations, interesting physical effects have been
predicted. For instance, for a nano-mechanical resonator coupled to a
microwave cavity,~\cite{rae,teufel,teufel2}
to field driven quantum dot~\cite{rabl} or in configurations with
double quantum dots,~\cite{zippilli,ouy} phonon cooling has been
found. With radio frequency quantum dots~\cite{ruskov} and a
electromagnetic cavity~\cite{clerk2} squeezing of the vibron position
and momentum quadratures has been theoretically predicted.\\

\noindent Even the simple AH model for a single electronic level
coupled to an undamped vibrational mode exhibits a rich physics, part
of which is still unexplored.  Roughly speaking, two regimes have been
considered so far, according to the ratio
\begin{equation}
\label{gamma}
\gamma=\frac{\Gamma_{0}}{\omega_{0}}
\end{equation}
between the vibron frequency, $\omega_{0}$, and the average dot-leads
electron tunneling rate, $\Gamma_{0}$.

\noindent For {\em fast} vibrations, $\gamma\ll 1$, every electron
tunneling event occurs over many oscillator periods. Then the
electrons are not sensitive to the position of the oscillator, but
only to its energy.~\cite{pisto,boubsi} Consequently, the oscillator
density matrix becomes close to diagonal in the basis of the energy
eigenstates.~\cite{boubsi} Many groups focused their attention on this
regime, employing rate equations to study transport phenomena as the
Franck-Condon blockade, super-Poissonian shot noise
~\cite{koch,mitra,shen,koch2} and even peculiar effects such as
sub-Poissonian out of equilibrium vibron
distributions.~\cite{rod2,merlo,cava3,brandesfano,brandesfano2}

\noindent In the opposite regime of {\em slow} vibrations $\gamma\gg
1$, electrons are extremely sensitive to the position of the
oscillator, which can be treated in a \textit{semi-classical}
approximation.~\cite{usmani,armour,armour2,nocera} Indeed, Mozyrsky
\textit{et al.} have shown~\cite{mozy} that it is the onset of a
semi-classical Langevin dynamics. In this regime the electronic
properties of the system have been especially investigated, in
particular the current and shot noise~\cite{mozy} in both limits of
weak~\cite{mozy2} and strong electron-vibron coupling.~\cite{pisto} In
the latter case, bi-stability and switching have been
addressed.~\cite{galp3,galp4,galp5} Recently, the classical phase
space of the vibron has also been studied.~\cite{huss}\\

\noindent Less attention has been devoted so far to the {\em coherent}
regime, where the {\em off-diagonal} elements of the system density
matrix in the energy representation play a relevant role. In this
regime, which starts around $\gamma\gtrsim 1$ (for a more precise
discussion, see Sec.~\ref{sec:regimes}), the competition between the
vibron and the electron time scales gives rise to a tough theoretical
problem. Most of the results obtained so far, concerning bi-stability
and phase space analysis, have been obtained stretching somehow the
validity range of semi-classical approaches,~\cite{huss,mozy,mozy2} in
the limit of very low temperatures, $\tau\ll 1$, where
\begin{equation}
\label{tau}
\tau=\frac{k_{\mathrm B}T}{\hbar\omega_{0}} \, ,
\end{equation}
here $k_{\mathrm B}$ is the Boltzmann constant and $T$ the environment
temperature. The interplay of electron and vibron time scales is
expected to strongly influence the dynamics of the vibron.  For
instance, for a NEMS based on a metallic dot in the weak coupling
regime, the damping effect of tunneling electrons on the vibron
dynamics is maximal when $\omega_{0}\approx\Gamma_{0}$.~\cite{rodr}
Similar mechanisms could play a role also in the simpler model of a
single level quantum dot.\\

\noindent Motivated by these considerations, in this article we
investigate the vibronic properties in the {\em coherent} regime
$\gamma\gtrsim 1$.  Here, because of the off-diagonal structure of the
system density matrix, a simple rate equation is no longer
justified.~\cite{blum} We derive a {\em generalized master
  equation}~\cite{brandes1,rod,flindt,nov,sch2} in the sequential
tunneling regime, in which all off-diagonal elements of the reduced
density matrix in the energy eigenbasis are retained. In the limit of
high temperatures considered here, $\tau>1$, a fairly large number of
basis states have to be included.  This fact leads to a serious
numerical challenge. Our calculation extends up to
$\tau\leq\tau_{\mathrm{max}}$, where $\tau_{\mathrm{max}}\approx
10$. We remark that, as a difference with previous
studies,~\cite{mozy2} our approach is not restricted to small
electron-vibron coupling.\\
\newline Here is a summary of our findings.
\noindent With the exception of a region $\gamma \to \tau$, in the
stationary regime the vibron state can be approximately described in
terms of an {\em effective thermal distribution}. In the coherent
regime, the effective temperature is {\em lower} than the
environmental temperature. Here, the role of coherences is crucial
despite subtle.  Non-vanishing off-diagonal elements of the vibron
density matrix in the eigenbasis, despite being very small compared
with diagonal elements, originate the peculiar effective thermal
re-distributions of the vibron occupation probabilities.\\
\noindent When $\gamma \simeq \tau$, the system exhibits deviations
from the above effective thermal state. The off-diagonal elements are
larger, leading to a marked suppression of the vibron fluctuations
even below the Poissonian value. The main results of our paper concern
the stationary vibron properties in the coherent regime and can be
summarized as follows: \\
\noindent ($i$) a {\em cooling} of the vibrational mode with respect
to the temperature of the electronic environment;\\ 
\noindent ($ii$) a strong {\em suppression} of the vibron Fano factor eventually
reaching {\em{sub-Poissonian}} values; \\
\noindent ($iii$) a reduction of the variances of the vibron position
and momentum quadratures.\\

\noindent We remark that the reported {\em cooling} phenomenon is a
direct consequence of the NEMS entering the {\em coherent} regime.  It
is \emph{not} ``induced'' by any external drive or dynamics, like
connecting the system to several reservoirs at different
temperatures.~\cite{pisto_c}\\
\noindent All the above effects are more pronounced when the electron-vibron
coupling strength is increased and none of them comes out treating the
system with a simple rate equation involving the diagonal matrix
elements only.\\

\noindent The paper is structured as follows. In Sec.~\ref{sec:model}
we describe the Anderson-Holstein model and the derivation of the
generalized master equation in the stationary regime. In
Sec.~\ref{sec:results} we illustrate the coherence effects on the
vibron behavior and on the electronic degree of freedom. The large
discrepancy with respect to the results obtained by means of a simple
rate equation is highlighted.  Conclusions are drawn in
Sec~\ref{sec:concl}.

\section{Model and Methods}
\label{sec:model}
\subsection{Anderson-Holstein Model}
\label{sec:AHmodel}
In the AH model,~\cite{flens,mitra,koch2} the hamiltonian
\begin{equation}
\label{eq:hlambda}
H=H_{\rm{dot}}+H_{\rm{osc}}+H_{\rm{int}}\,,
\end{equation}
describes a ultra-small quantum dot ($H_{\rm{dot}}$) coupled to an
harmonic oscillator ($H_{\rm{osc}}$) via the coupling term
$H_{\rm{int}}$.  The quantum dot is modeled~\cite{koenig96} as a spin
degenerate single level with the average level spacing of the order of
the charging energy $E_{\rm{C}}$ (from now on, $\hbar=1$)
\begin{equation}
\label{eq:model1}
H_{\rm{dot}}=\epsilon \hat n+E_{\mathrm{C}} \hat n(\hat n-1).
\end{equation}
Here, $\hat n=\sum_{\sigma=\pm1} \hat n_{\sigma}$ is the occupation
number of the level, with $\hat
n_{\sigma}=d^{\dag}_{\sigma}d_{\sigma}$ the occupation of spin
$\sigma/2$ with $\sigma=\pm1$ and $d_{\sigma}$, $d^{\dag}_{\sigma}$
are the fermionic dot operators.  We assume that $E_{\mathrm{C}}$ is
the largest energy scale of the problem and we consider only single
excess occupancy on the dot, $n=0,1$.  The energy
$\epsilon=\xi+2E_{\rm{C}}(1/2-n_{\rm{g}})$ includes the energy of the
lowest unoccupied single-particle level $\xi$ and a term connected to
$n_{\rm{g}}=C_{\rm{g}}V_{\rm{g}}/e$, the charge induced by the gate
voltage $V_{\rm{g}}$ with gate capacitance $C_{\rm{g}}$ ($-e$ is the
electron charge).~\cite{cava3}

The vibron is described as an harmonic oscillator with mass $m$ and
frequency $\omega_0$.  In terms of the boson operators $b$, $b^{\dag}$
it is modeled as
\begin{equation}
H_{\rm{osc}}=\omega_0\left(b^{\dag}b+1/2\right).
\end{equation}

\noindent The dot and the oscillator are coupled via a term bi-linear
in the oscillator position $x$ and in the effective charge number on
the dot, $\hat n-n_{\rm{g}}$,~\cite{zazunov,shen}
\begin{equation}
\label{eq:coupsbm}
H_{\rm{int}}=\sqrt{2}\lambda\omega_{0}\frac{x}{\ell_0}(\hat n-n_{\rm{g}})\,,
\end{equation}
where $\lambda$ is the adimensional coupling parameter and 
\begin{equation}
\label{eq:ell0}
\ell_{0}=\frac{1}{\sqrt{m\omega_{0}}}
\end{equation}
is the characteristic length of the harmonic oscillator.

\noindent The dot is coupled also to the external left ($\rm{L}$) and
right ($\rm{R}$) leads of non interacting electrons
\begin{equation}
\label{eq:sbm}
H_{\rm{leads}}=\!\!\!\!\sum_{\alpha=\rm{L},\rm{R}}\sum_{k,\sigma=\pm1}\varepsilon_{k}\ c^{\dagger}_{\alpha,k,\sigma}c_{\alpha,k,\sigma}\, ,
\end{equation}
where $c_{\alpha,k,\sigma}$ and $c^{\dagger}_{\alpha,k,\sigma}$ are
the fermionic operators.  The leads are assumed in equilibrium with
respect to their electrochemical potential
$\mu_{\rm{L},\rm{R}}=\mu_{0}\pm eV/2$, where $V$ is a symmetrically
applied bias voltage, and $\mu_{0}$ is the reference chemical
potential.  The bias $V$ forces electrons to flow from the left to the
right lead through the dot via the tunneling hamiltonian
\begin{equation}
\label{eq:tunnel}
H_{\rm{t}}=t_0\sum_{\alpha=\rm{L},\rm{R}}\sum_{k,\sigma=\pm1}c^{\dagger}_{\alpha,k,\sigma}d_{\sigma}+\rm{h.c.}\, ,
\end{equation}
where $t_{0}$ is the tunneling amplitude 
through both the left  and right barriers.

\noindent Equation~(\ref{eq:hlambda}) can be diagonalized by the
Lang--Firsov polaron transformation,~\cite{zazunov} with generator
\begin{equation}
\label{eq:platrafo}
\mathcal
U=\exp{[\hat \eta \, (b^{\dagger}-b)]}\quad\mbox{and}\quad \hat \eta=\lambda \, (\hat n-n_{\rm{g}}).
\end{equation}
This procedure imposes no restriction on the possible values of
$\lambda$.  The transformed operators in the polaron frame
$\bar{\mathcal{O}}=\mathcal{U}\mathcal{O}\mathcal{U}^{\dagger}$ are
$\bar{b}=b- \hat \eta$ and
$\bar{d}_{\sigma}=d_{\sigma}\exp{\left[\lambda(b-b^{\dagger})\right]}$,
while $\hat n$ is invariant.  The diagonal Hamiltonian, expressed in
terms of the original operators reads
\begin{equation}
\label{eq:model3}
\bar{H}=\bar{\epsilon}  \, \hat n + \omega_{0}(b^{\dag}b+1/2)\ ,
\end{equation}
with renormalized level position
$\bar{\epsilon}=\xi+(E_{\mathrm{C}}-\lambda^2\omega_0)(1-2n_{\mathrm{g}})$.
In the following we choose $\mu_0=\xi$ setting the resonance between
the $n=0,1$ states at $n_{\rm{g}}=1/2$. The eigenstates of
Eq.~(\ref{eq:model3}) will be denoted as $|n,l\rangle$, where $n$ is
the dot occupation number and $l$ represents the vibron number.
\noindent The transformed tunneling Hamiltonian is
\begin{equation}
\label{eq:tunnel2}
\bar{H}_{\rm{t}}=t_{0}\sum_{\alpha=\rm{L},\rm{R}}\sum_{k,\sigma=\pm1}e^{\lambda(b-b^{\dagger})}c^{\dagger}_{\alpha,k,\sigma}d_{\sigma}+\rm{h.c.}\,.
\end{equation}

\subsection{Master Equation}
The dynamics of the dot and the oscillator is described by the reduced
density matrix ${\bar{\rho}}(t)$, defined as the trace over the leads
of the total density matrix $\bar{\rho}_{\rm{tot}}(t)$, in the polaron
frame
\begin{equation}
\bar{\rho}(t)=\mathrm{Tr}_{\rm{leads}}\{{\bar{\rho}}_{\rm{tot}}(t)\}\,.
\end{equation}
We consider the sequential, weak tunneling regime, treating
$\bar{H}_{\rm{t}}$ in Eq.~(\ref{eq:tunnel2}) to the lowest order.
This approximation is valid for not too low temperatures $T$,
$\Gamma_0=2\pi|t_0|^2\nu<k_{\rm{B}}T$, where $\nu$ is the leads
density of states.  We further perform the Born
approximation,~\cite{timm} assuming that the system and the leads are
independent before $H_{\mathrm{t}}$ is switched on.  This amounts to
take the factorized form at $t=0$,
$\bar{\rho}_{\rm{tot}}(0)=\bar{\rho}(0)\otimes\rho_{\rm{l}}(0)$, where
$\rho_{\rm{l}}(0)=\rho_{\rm{L}}(0)\otimes\rho_{\rm{R}}(0)$ and
$\rho_{\rm{L/R}}(0)$ are the initial equilibrium density matrices of
the left and the right lead respectively.

In the interaction picture with respect to $\bar{H}_{\rm{t}}$,
any operator, $A$, is transformed as
\begin{equation}
\nonumber
A_{\rm{I}}=e^{i(\bar{H}+H_{\rm{leads}})t}Ae^{-i(\bar{H}+H_{\rm{leads}})t}.
\end{equation}

\noindent The reduced density matrix evolves according to
\begin{equation}
\begin{split}
\label{eq:master3}
\!\!\!\!\!\dot{\bar{\rho}}_{\rm{I}}(t)&\!\!=\!-\!\!\!\!\sum_{\sigma=\pm1}\!\int_{0}^{t}\!\!{\rm
  d}t'\!\left\{\![Q_{\rm{I},\sigma}(t),Q^{\dagger}_{\rm{I},\sigma}(t')\bar{\rho}_{\mathrm{I}}(t')]K^{+}(t-t')\right.\\
&-[Q_{\rm{I},\sigma}(t),\bar{\rho}_{\rm{I}}(t')Q^{\dagger}_{\rm{I},\sigma}(t')]K^{-}(t'-t)\\
&+[Q^{\dagger}_{\rm{I},\sigma}(t),Q_{\rm{I},\sigma}(t')\bar{\rho}_{\rm{I}}(t')]K^{-}(t-t')\\
&-\left.[Q^{\dagger}_{\rm{I},\sigma}(t),\bar{\rho}_{\rm{I}}(t')Q_{\rm{I},\sigma}(t')]K^{+}(t'-t)\right\}\, .
\end{split}
\end{equation}
Here,
\begin{equation}
\nonumber
Q_{\rm{I},\sigma}(t)=e^{\lambda(b_{\rm{I}}(t)-b_{\rm{I}}^{\dagger}(t))}d_{\rm{I},\sigma}(t)\,,
\end{equation}
and $K^{\pm}(t)=K_{\mathrm L}^{\pm}(t)+K_{\mathrm R}^{\pm}(t)$ are the
leads correlation functions
\begin{equation}
\begin{split}
\label{eq:correlations}
K_{\alpha}^{+}(t)
&=|t_0|^2\sum_{k}\mathrm{Tr}_{\mathrm{leads}}\left\{c^{\dag}_{\alpha,k,\sigma}(t)c_{\alpha,k,\sigma}(0)\rho_{\rm{l}}\right\},\\
K_{\alpha}^{-}(t)
&=|t_0|^2\sum_{k}\mathrm{Tr}_{\mathrm{leads}}\left\{c_{\alpha,k,\sigma}(t)c_{\alpha,k,\sigma}^{\dag}(0)\rho_{\rm{l}}\right\}\,.\\
\end{split}
\end{equation}
In Eq.~(\ref{eq:master3}) we performed the large reservoirs approximation~\cite{timm}
\begin{equation}
\bar{\rho}_{\mathrm{I,tot}}(t')=\bar{\rho}_{\mathrm{I}}(t')\cdot\rho_{l}
\end{equation}
assuming that tunneling events have a negligible effect on the leads,
which remain in the thermal equilibrium state, denoted as
$\rho_{\rm{l}}$. In the weak tunneling regime ($\Gamma_0<k_{\rm{B}}T$)
one can replace $\bar{\rho}_{\mathrm{I}}(t')\approx\bar{\rho}_{\mathrm
  I}(t)$ using the standard Markov approximation, and extend the
integration limit to $\infty$.

Eq.~(\ref{eq:master3}) can be projected on the eingenstates of the
hamiltonian~(\ref{eq:model3}), obtaining an infinite set of coupled
equations for the density matrix elements $\langle
n,l|\bar{\rho}(t)|n',l'\rangle$ (where $n,n'\in\{0,1\}$ and $l,l'\geq
0$).  It can be easily shown that diagonal and off-diagonal elements
in the \textit{electron} number decouple and in the stationary regime
the latter tend to zero. In fact, the coupling of the leads and the dot charge
leads to a rapid decay of superpositions of dot states with
different charges.~\cite{gurv,timm} Since we are interested in the
stationary properties, we focus on the density matrix elements
diagonal in the \textit{electron} level occupation number,
$\bar{\rho}_{qq'}^{n}(t)=\langle
n,q|\bar{\rho}(t)|n,q'\rangle\nonumber$.  They obey the following
generalized master equation (GME)
\begin{equation}
\label{eq:interaz}
\dot{\bar{\rho}}_{\mathrm{I},qq'}^{n}(t) = \sum_{pp'n'}\mathcal{R}_{qq'pp'}^{nn'}
e^{i\omega_0(q-q'-p+p')t} \; \, \bar\rho^{n'}_{\mathrm{I},pp'}(t)\,,
\end{equation}
where $\mathcal{R}_{qq'pp'}^{nn'}$ are the Redfield tensor elements~\cite{blum}
\begin{equation}
\begin{split}
\label{eq:redfield1}
\mathcal{R}_{qq'pp'}^{01}=&X_{qp}X_{q'p'}\left[C^-(\omega_{pq})^*+C^-(\omega_{p'q'})\right]\, ,\\
\mathcal{R}_{qq'pp'}^{11}=&-\sum_{l}\left[X_{lq'}X_{lp'}C^-(\omega_{p'l})\delta_{pq}\right.\\
 &\left.+X_{lq}X_{lp}C^-(\omega_{pl})^*\delta_{p'q'}\right]\,,
\end{split}
\end{equation}
\begin{equation}
\begin{split}
\label{eq:redfield2}
\mathcal{R}_{qq'pp'}^{10}=&2X_{pq}X_{p'q'}\left[C^+(\omega_{qp})+C^+(\omega_{q'p'})^*\right]\, ,\\
\mathcal{R}_{qq'pp'}^{00}=&-2\sum_{l}\left[X_{q'l}X_{p'l}C^+(\omega_{lp'})^*\delta_{pq}\right.\\
 &\left.+X_{ql}X_{pl}C^+(\omega_{lp})\delta_{p'q'}\right]\, .
\end{split}
\end{equation}
Here $\omega_{qq'}\equiv\omega_0(q-q')$ and 
\begin{equation}
\begin{split}
\label{franck}
X_{qq'}&=\langle q|e^{\lambda(b-b^{\dag})}|q'\rangle\\
&=e^{-\frac{\lambda^2}{2}}\sqrt{\frac{q_<!}{q_>!}}
L^{|q'-q|}_{q_<}(\lambda^2)\left[\mathrm{sgn}(q'-q)\lambda\right]^{|q'-q|}\,,\\
\end{split}
\end{equation}
are the generalized Franck-Condon factors~\cite{koch,cava3}, with
$q_<=\min\left\{q,q'\right\}$, $q_>=\max\left\{q,q'\right\}$ and
$L^n_q(x)$ the generalized Laguerre polynomials.~\cite{abra} The
factors 2 in Eqs. (\ref{eq:redfield2}) are due to the spin degeneracy.
In Eqs. (\ref{eq:redfield1}),(\ref{eq:redfield2}) the generalized
tunneling rates $C^{\pm}(\omega_{qq'})=C_{\mathrm
  L}^{\pm}(\omega_{qq'})+C_{\mathrm R}^{\pm}(\omega_{qq'})$ have also
been introduced
\begin{equation}
C^{\pm}_{\alpha}(\omega_{qq'})=\int_0^{\infty}\mathrm{d}\theta K^{\pm}_{\alpha}(\pm\theta)e^{-i(\bar{\epsilon}+\omega_{qq'})\theta}\,.
\end{equation}

\noindent Exploiting the identity
\begin{equation}
\nonumber
\int_0^{\infty}\mathrm{d}\theta\exp{\left(i\Omega \theta\right)}=\pi\delta(\Omega)+iP.V.(1/\Omega)\, ,
\end{equation}
and the explicit form of the leads correlation function~\cite{scholler}
\begin{equation}
K^{\pm}_{\alpha}(t)
={-}\frac{i\Gamma_{0}e^{\pm i\mu_{\alpha}t}}{2\beta \sinh{[\frac{\pi}{\beta}(t-\frac{i}{\omega_{\rm{c}}})]}}\,,\nonumber\\
\end{equation}
with $\omega_{\rm{c}}$ the cut-off energy and $\beta=1/(k_{\rm{B}}T)$, one has
\begin{equation}
\label{eq:rate_c}
C^{\pm}(\omega_{qq'})=\frac{\Gamma_0}{2}\sum_{\alpha}\left[f_{\alpha}^{\pm}(\omega_{qq'})\pm i\Delta_{\alpha}(\omega_{qq'})\right]\,.
\end{equation}
Here
\begin{equation}
f_{\alpha}^{+}(E)=\frac{1}{1+e^{\beta(\bar{\epsilon}+E-\mu_{\alpha})}}
\end{equation}
is a Fermi function, $f^-_{\alpha}(E)=1-f^+_{\alpha}(E)$ and
\begin{equation}
\Delta_{\alpha}(\omega_{qq'})=\log\left(\frac{2\pi}{\beta\omega_{\mathrm{c}}}\right)+\mathrm{Re}\,\psi\left[\frac{1}{2}+\frac{i\beta}{2\pi}(\bar{\epsilon}+\omega_{qq'}-\mu_{\alpha})\right]\nonumber
\end{equation}
the depolarization shift. The stationary system properties are obtained
from the solution of the steady-state GME
\begin{equation}
\label{eq:staz}
\sum_{pp'n'}\left[\mathcal{R}_{qq'pp'}^{nn'}+i\omega_0\left(p'-p\right)\delta_{pq}\delta_{p'q'}\right]\bar\rho^{n'}_{pp'}=0\,,
\end{equation}
written here in the Schr\"odinger representation, for the stationary
values of the reduced density matrix elements
\begin{equation}
\bar{\rho}_{pp'}^{n}=\lim_{t\to\infty}\bar\rho_{pp'}^n(t)\,.
\end{equation}
Note that the GME~(\ref{eq:staz}) takes into account both {\em all}
off-diagonal (coherences) and diagonal terms in the \textit{vibron
  number}.\\
\noindent The steady state current calculated e.g. on the right tunnel
barrier, reads
\begin{eqnarray}
\label{eq:currentGME}
I=e\Gamma_{0}&&\sum_{pqq'}\left\{-2\left[C^{+}_{\mathrm{R}}(\omega_{pq'})+C^{+}_{\mathrm{R}}(\omega_{pq})^*\right]X_{qp}X_{q'p}\bar{\rho}_{qq'}^{0}\right.\nonumber\\
&&+\left.\left[C^{-}_{\mathrm{R}}(\omega_{qp})+C^{-}_{\mathrm{R}}(\omega_{q'p})^*\right]X_{pq}X_{pq'}\bar{\rho}_{qq'}^{1}\right\}\, .
\end{eqnarray}
Note that in the steady-state the current is independent of the
barrier index.
\subsection{Rotating Wave Approximation}
\label{sec:RWA}
In the regime of fast vibrational motion, $\gamma \ll 1$, the
contribution of fast oscillatory terms to the solution of
Eq.~(\ref{eq:interaz}) is negligible.  It is then possible to perform
the rotating wave approximation (RWA), neglecting the oscillating
contributions and including only the dominant secular
terms.~\cite{blum} They are those which connect density matrix
elements with $p-p'=q-q'$.  This implies that the diagonal elements
$\bar{\rho}_{qq}^{n}$ are decoupled from the off-diagonal ones
$\bar{\rho}_{pp'}^{n}$ with $p'\neq p$.  In addition, the non-diagonal
elements vanish in the stationary regime.~\cite{mitra} Hence
stationary properties are fully described by the diagonal occupation
probabilities $\bar{P}_{nq}=\bar{\rho}^n_{qq}$ and Eq.~(\ref{eq:staz})
reduces to a standard rate equation
\begin{equation}
\label{eq:RWA}
z_n\bar{P}_{nq}\sum_{n'\neq n}\sum_{p=0}^{\infty}\Gamma_{qp}^{nn'}-\sum_{n'\neq n}\sum_{p=0}^{\infty}z_{n'}\bar{P}_{n'p}\Gamma_{pq}^{n'n}=0\, .
\end{equation}
The coefficients $z_n$ stem from the spin degeneracy: $z_0=2$,
$z_1=1$. The tunneling rates
$\Gamma_{pq}^{nn'}=\Gamma_{{\mathrm{L}},pq}^{nn'}+\Gamma_{\mathrm{R},pq}^{nn'}$
for the transition $|n, p\rangle\rightarrow|n', q\rangle$, are
\begin{equation}
\begin{split}
\label{eq:rate_RWA}
\!\!\!\!\Gamma_{\alpha,qp}^{nn+1}=&\,2X_{pq}^2\mathrm{Re}\left\{C_{\alpha}^+(\omega_{pq})\right\}\!=\!\Gamma_0 X_{pq}^2 f^+_{\alpha}(\omega_{pq});\\
\!\!\!\!\Gamma_{\alpha,pq}^{n+1n}=&\,2X_{pq}^2\mathrm{Re}\left\{C_{\alpha}^-(\omega_{pq})\right\}\!=\!\Gamma_0 X_{pq}^2 f^-_{\alpha}(\omega_{pq}).
\end{split}
\end{equation}
Note that Eq.~(\ref{eq:RWA}) does not depend on $\gamma$.\\
\noindent Within the RWA, the steady-state current
Eq.~(\ref{eq:currentGME}) is
\begin{equation}
\label{eq:currentRWA}
I_{\rm{RWA}}=e\Gamma_{0}\sum_{p,q}\left[-2\bar{P}_{0q}\Gamma_{\mathrm{R},qp}^{0,1}+\bar{P}_{1q}\Gamma_{\mathrm{R},qp}^{1,0}\right]\, .
\end{equation}
\begin{figure}[t!]
\begin{center}
\includegraphics[width=7cm,keepaspectratio]{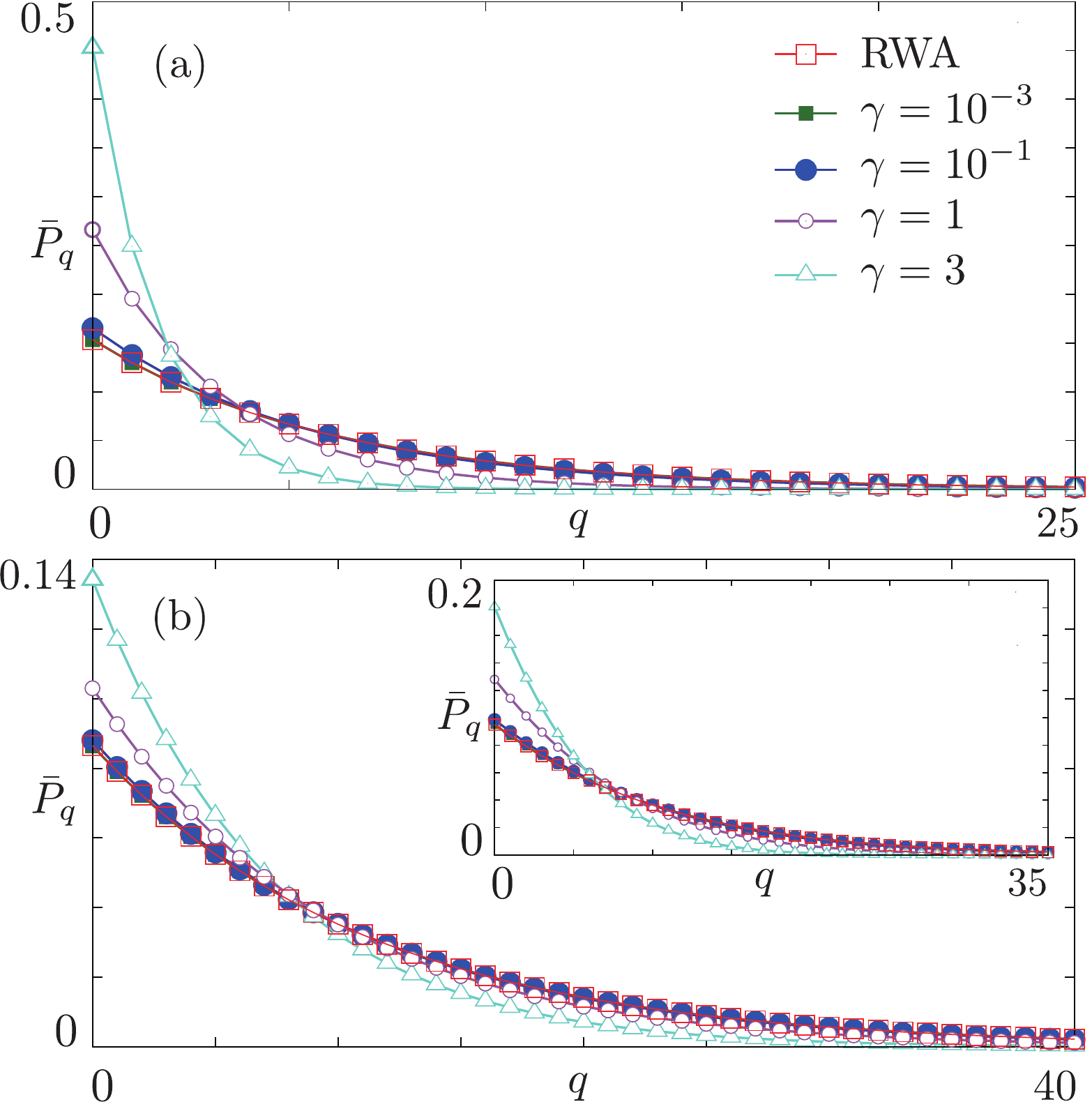}
\caption{(Color online) Vibron occupation probabilities $\bar{P}_q$ as
  a function of the vibron number $q$, calculated numerically solving
  the RWA rate equation in Eq.~(\ref{eq:RWA}) (red empty squares) and
  the GME in Eq.~(\ref{eq:staz}). Here $\gamma=3$ (cyan empty
  triangles), $\gamma=1$ (purple empty circles), $\gamma=10^{-1}$
  (blue circles), $\gamma=10^{-3}$ (green squares). (a) $\lambda=0.5$
  and $eV=\omega_{0}$. (b) Main panel: $\lambda=0.5$ and
  $eV=20\,\omega_0$. Inset: $\lambda=2$ and $eV=20\,\omega_{0}$.  In
  all panels other parameters are $\tau=6$, $n_{\rm{g}}=1/2$ and
  $\omega_{\rm{c}}=10^{6}\,\omega_0$.}
\label{fig:fig3}
\end{center}
\end{figure}
Results presented in the following Section are obtained by numerical
solution of the GME Eq.~(\ref{eq:staz}) truncating the size of the
vibron Hilbert space until convergence is reached (see next Section
for details).  In the limit $\gamma \ll 1$, this approach accurately
reproduces the well-known solution of the RWA rate
equation.~\cite{rodr,pisto_c,merlo,koch,koch2}

\noindent This is illustrated in Fig.~\ref{fig:fig3}, where the
diagonal elements of the reduced density matrix,
$\bar{P}_{q}=\sum_{n}\bar{\rho}^n_{qq}$, obtained by solving the GME
for different values of $\gamma$ are shown. The solutions of the GME
converge to those in the RWA for $\gamma\rightarrow0$ (red squares),
both for small and large voltages (panels (a)~-~(b)) and even in the
strong electron-vibron coupling regime (panel c).  This convergence
has been systematically observed in the whole range of parameters and
constitutes a validation of our numerical procedure.\\
\noindent On the other hand, with increasing $\gamma$, deviations from
the RWA are obtained. They are originated from the
coherent off-diagonal elements of $\bar{\rho}_{qq'}$.  These
deviations represent the central part of our work and relevant consequences 
will be discussed in details in the following Section.

\section{Results}
\label{sec:results}
In the present Section, we investigate the stationary equilibrium
properties of the system as a function of the parameters $\gamma$,
$\tau$ defined in Eqs.(\ref{gamma}), (\ref{tau}). Here, $\gamma$
distinguishes between fast, slow or coherent regimes, and $\tau$ is
the reduced temperature. We focus mainly on the vibronic
properties. Relevant quantities of our interest are the vibron Fano
factor and the position and momentum quadratures. Their behavior can
be understood by first analyzing the structure of the oscillator
density matrix in the vibron eigenbasis. In the final part of this
Section, the electronic current and the average electronic occupation
of the dot will be briefly addressed.

\subsection{Parameters regimes}
\label{sec:regimes}
\begin{figure}[t!]
\begin{center}
\includegraphics[width=8cm,keepaspectratio]{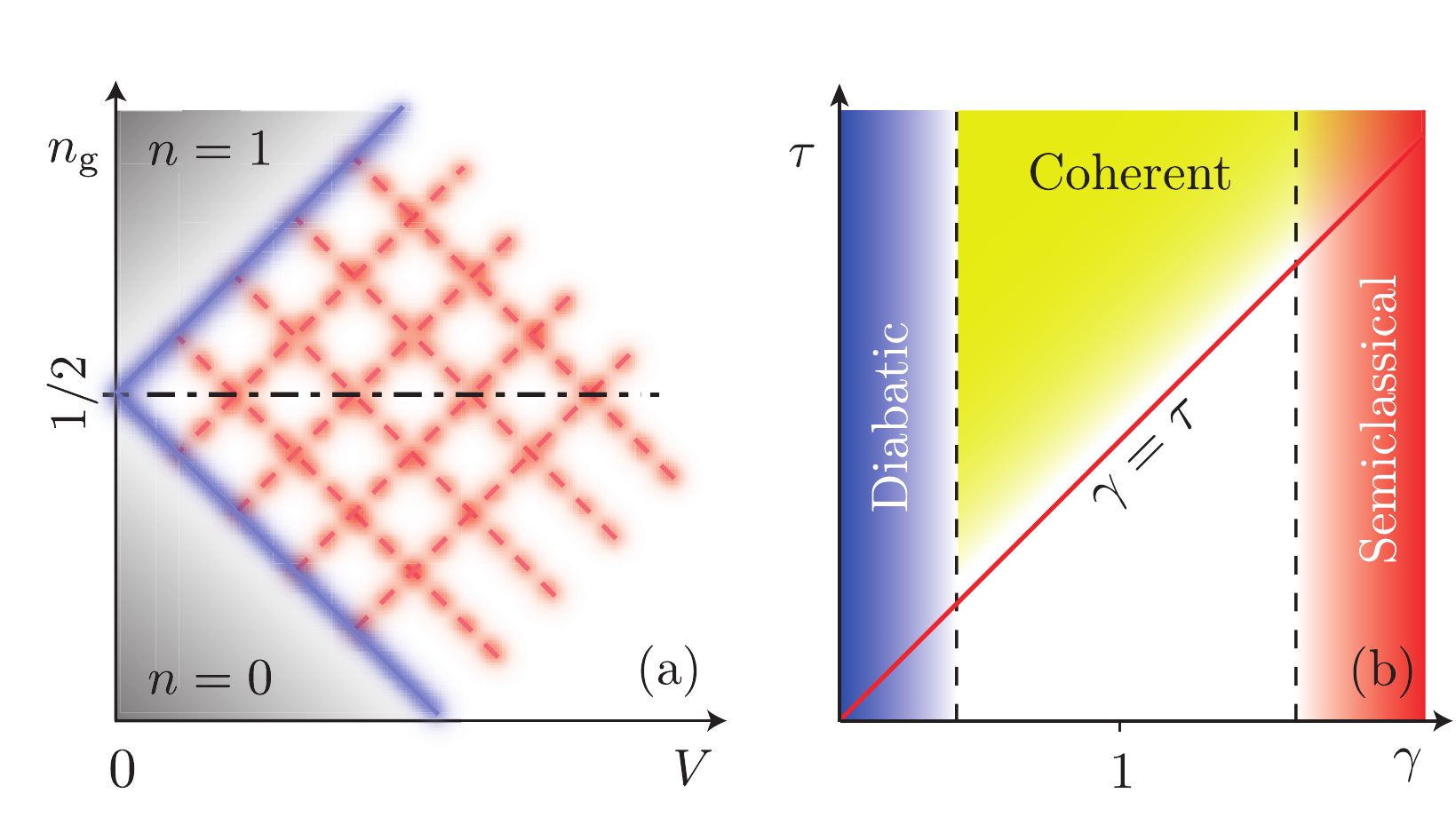}
\caption{(Color online) (a) Stability diagram in the $(V,n_{\mathrm
    g})$ plane: gray shaded regions denotes Coulomb Blockade
  regime. Blue (solid) lines mark the onset of transitions between
  ground states, red (dashed) lines signal transitions involving
  excited states of the vibron. (b) Different regimes as a function of
  $\gamma=\Gamma_{0}/\omega_{0}$ and $\tau=k_{\mathrm
    B}T/\omega_{0}$. The blue shaded region denotes the ``diabatic''
  regime of fast vibrations, the red one the semi-classical regime
  with $\omega_{0}\ll\Gamma_{0}$. The region above the line
  $\gamma=\tau$ marks the region accessible by our methods and the
  yellow shaded area indicates the accessible regime where coherences
  among vibron states become important. }
\label{fig:fig1}
\end{center}
\end{figure}
Figure~\ref{fig:fig1}(a) represents the stability diagram of the
system as a function of the bias voltage $V$ and of $n_{\mathrm
  g}$. In the shaded (gray) regions the system is in the Coulomb
blockade regime with the dot empty (full) if $n_{\mathrm g}<1/2$
($n_{\mathrm g}>1/2$). The blue (solid) lines mark transitions between
the states $|0,q\rangle$ and $|1,q\rangle$ ($n$ and $q$ are the
occupation number of the electronic or vibronic states
$|n,q\rangle$). Red (dashed) lines indicate the activation threshold
for transport channels with transitions $|n,q\rangle\to|n',q'\rangle$
where in general $q,q'\neq 0$. The boundaries of different transport
regions are thermally broadened.\\
\noindent In the following analysis we will focus at $n_{\mathrm
  g}=1/2$ where the $n=0$ and $n=1$ charge states are on resonance
(dashed-dotted line in Fig.~\ref{fig:fig1}(a)). Qualitatively similar
results are observed also off-resonance at $n_{\mathrm g}\neq 1/2$.\\
\noindent In Fig.~\ref{fig:fig1}(b) we report a sketch of the system
behavior in the $(\gamma,\tau)$ plane.\\

We distinguish a ``{\em diabatic}'' regime of fast vibrations for
$\gamma\ll 1$ (blue shaded region), where the RWA is
valid,~\cite{rodr,pisto_c,merlo,koch,koch2} and an ``{\em adiabatic}''
regime of slow vibrations $\gamma\gg 1$ where semiclassical methods
have been applied~\cite{huss,nocera,pisto,galp4,mozy} with approaches
confined to low temperatures $\tau\ll 1$.\\
\noindent The regime in between these two regions is, on the other hand,
unexplored. Our GME method fills in precisely this gap. Indeed, we will
show that, increasing $\gamma$ from the diabatic regime, marked
deviations from the solution of the RWA rate equations appear. Here, coherences represented by the
off-diagonal elements of the reduced density matrix  are
relevant. The constraints on our GME method are:\\
\noindent ($i$) the sequential tunneling approximation which implies
$0<\gamma<\tau$ (the area above the red line in Fig.~\ref{fig:fig1});\\
\noindent ($ii$) a restriction on the temperature
$\tau<\tau_{\mathrm{max}}\approx 10$, due to the rapid increase with
$\tau$ of the number of vibron states needed for the convergence of
our numerical calculations.\\
\noindent Because of these constraints, the 
transition towards the semiclassical limit cannot be explored.\\

\noindent We solve numerically the GME in the steady state, increasing
the size of the vibron Hilbert space including up to 150 oscillator
states per charge state in the basis, until convergence is reached.
The limitation on the number of oscillator states employed in the
calculation is essentially due to computer memory requirements and by
the decreasing rate of convergence. At high temperature
$\tau\approx10$, already more than 100 oscillator basis states are
required for convergence.
\subsection{Overview of the results}
\label{sec:overview}
\begin{figure}[t!]
\begin{center}
\includegraphics[width=6cm,keepaspectratio]{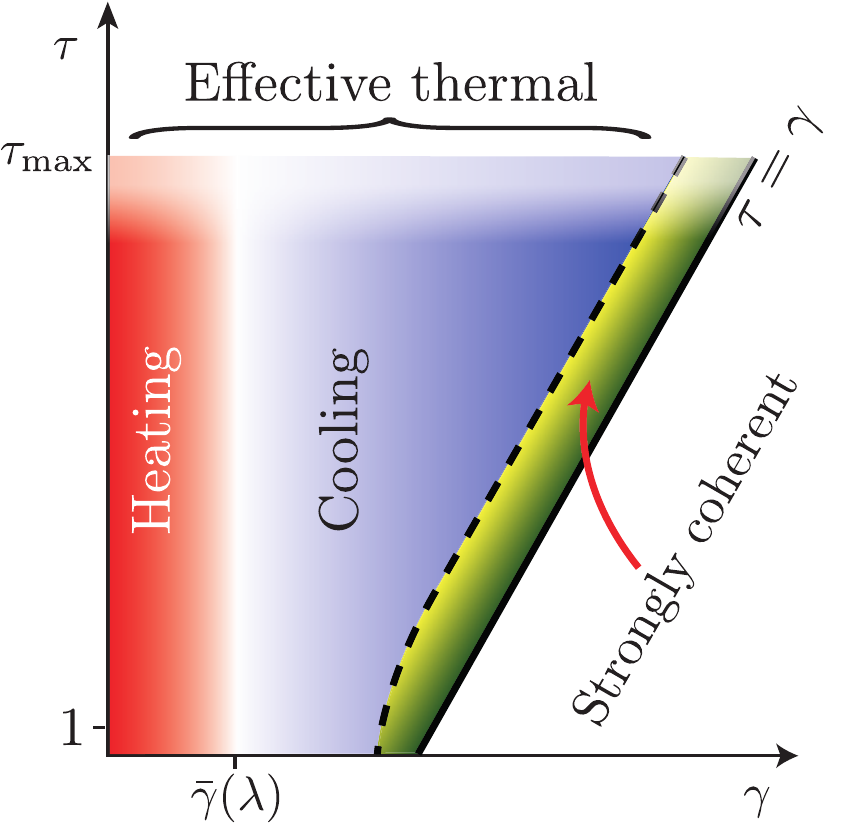}
\caption{(Color online) Schematic overview of the results in the
  $(\gamma,\tau)$ plane, see text for details on the notation.}
\label{fig:fig2}
\end{center}
\end{figure}
Before entering our analysis, we summarise in Fig.~\ref{fig:fig2} the
stationary vibronic characteristics we obtain in the various parameter
regimes. We address the temperature regime $1\lesssim\tau\lesssim
\tau_{\mathrm{max}}$ and $0 < \gamma \lesssim \tau$. We will show that
in a large parameters range (shaded area) the vibron density matrix
\begin{equation}
\bar{\rho}_{qq'}=\sum_{n=0,1}\bar{\rho}_{qq'}^{n}
\end{equation}
is well approximated by an {\em effective} thermal distribution with
temperature $\tau_{\rm{eff}}$
\begin{equation}
\label{eq:effective}
\bar{\rho}_{qq'}^{({\mathrm{th}})}(\tau_{\mathrm{eff}})=\left(1-e^{-1/\tau_{\mathrm{eff}}}\right)e^{-q/\tau_{\mathrm{eff}}}\delta_{q,q'}\, .
\end{equation}
For $\gamma\ll 1$ the off-diagonal elements of the reduced density
matrix are negligible and a fit of $\bar{\rho}_{qq'}$ to
Eq.~(\ref{eq:effective}) leads to $\tau_{\mathrm{eff}}\geq\tau$.  This
"heating" phenomenon is due to the finite voltage, and for $V\to 0$
the system attains a thermal equilibrium distribution in the polaron
frame, $\tau_{\mathrm{eff}}\to\tau$, as reported in
Ref.~\onlinecite{merlo}.\\
\noindent With increasing $\gamma$ the off-diagonal elements of
$\bar{\rho}_{qq'}$ increase. The onset of this coherent regime is
marked by the condition $\gamma\gtrsim\bar{\gamma}(\lambda)$, with the
latter a $\lambda$-dependent threshold value. For typical values
$\lambda\approx 1$ we find $0.1\leq\bar{\gamma}(\lambda)\leq 1$.  When
$\bar{\gamma}(\lambda) \lesssim \gamma < \tau$ the coherences are much
smaller than the diagonal elements of the reduced density matrix. In
this {\em weakly coherent} regime the vibron density matrix can still
be fitted with a thermal distribution, but at a lower effective
temperature, eventually reaching the {\em cooling} regime where
$\tau_{\mathrm{eff}}<\tau$. The cooling is always accompanied by a
reduction both of the fluctuations of the vibronic population and of
the variance of position and momentum quadratures.\\
\noindent A completely different system behavior takes place when
$\gamma\to\tau$, (yellow-green area delimited by the dashed and
continuous lines in Fig.~\ref{fig:fig2}). In this {\em strongly
  coherent} regime off-diagonal terms of the density matrix are
paramount and the diagonal part of the density matrix deviates from
the simple thermal distribution. Here, despite of the high temperature
($\tau>1$), a non-classical system behavior comes out.  Suppression of
the vibron populations fluctuations and of the variances of the
position and momentum quadratures persists also in this regime.

\subsection{Density matrix and cooling}
\begin{figure}[ht]
\begin{center}
\includegraphics[width=8cm,keepaspectratio]{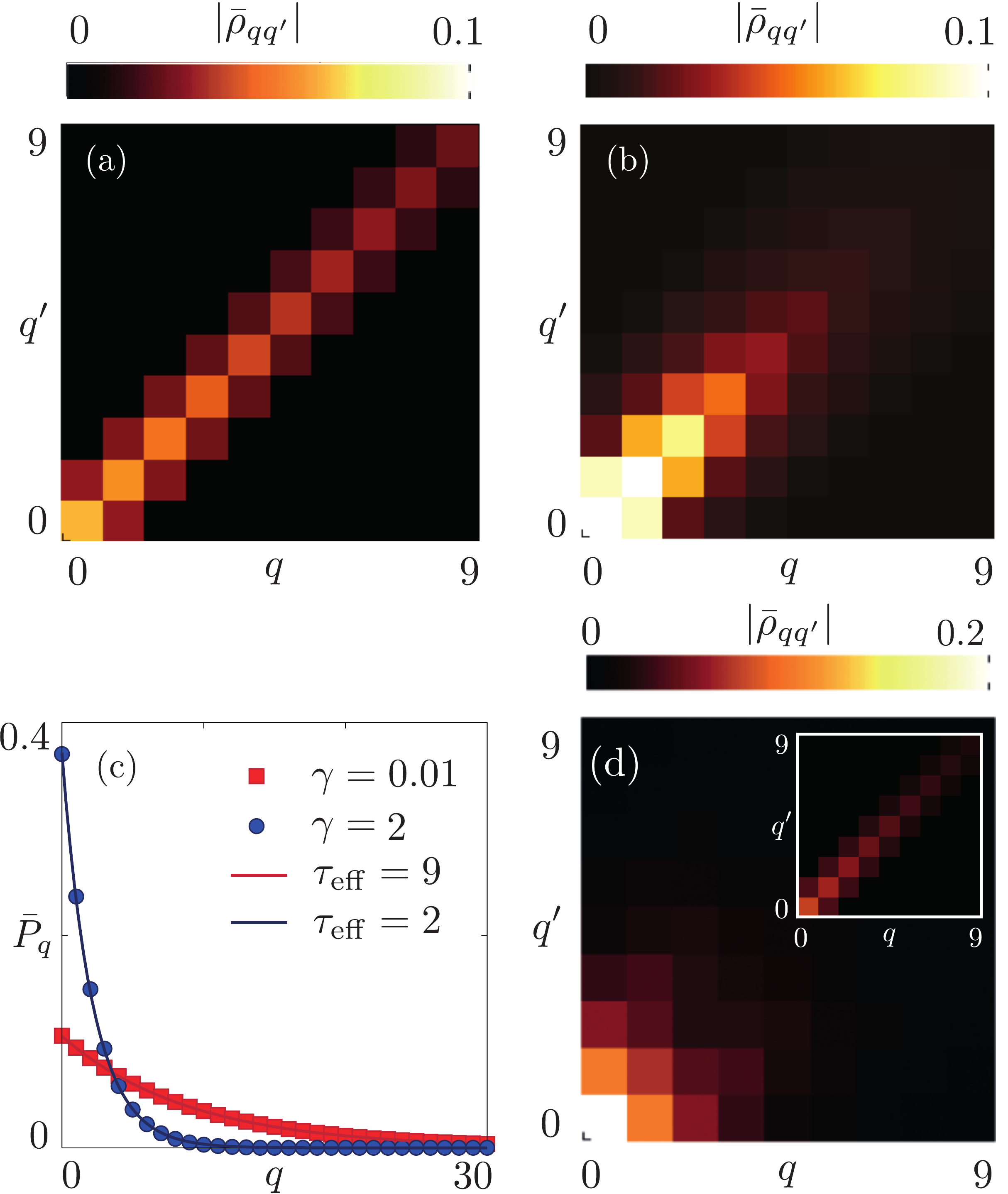}
\caption{(Color online) Density plot of $|\bar{\rho}_{qq'}|$ as a
  function of $q,q'$.  Panel (a): $\gamma=0.01$ and $\tau=9$; Panel
  (b): $\gamma=2$ and $\tau=9$.  
  Panel (c): Occupation
  probabilities $\bar{P}_{q}=\bar{\rho}_{qq}$ for $\tau=9$ and
  $\gamma=0.01$ (squares) or $\gamma=2$ (dots). Lines are fit to a
  thermal distribution with effective temperature
  $\tau_{\mathrm{eff}}$.
  In panel (d) $\gamma=1.2$ and
  $\tau=3$ (main), and $\tau=9$ (inset).   In all panels, $n_{\rm{g}}=1/2$,
  $eV=\omega_{0}$, $\lambda=2$ and $\omega_c=10^6\ \omega_{0}$.}
\label{fig:fig4}
\end{center}
\end{figure}

We start our analysis investigating the structure of the oscillator
density matrix in the vibron eigenbasis in the regimes indicated in
Fig.~\ref{fig:fig2}.\\
\noindent We first concentrate on the density matrix $\bar{\rho}_{qq'}$ when it
is well described by the
effective thermal distribution, Eq.~(\ref{eq:effective}). In
Figs.~\ref{fig:fig4}(a)~-~(b) we report the density plots of $|\bar{\rho}_{qq'}|$ in
the diabatic and in the coherent regime respectively.
\noindent For $\gamma\ll1$ - panel (a) - the density matrix is
strongly peaked around the diagonal, $q\approx q'$. The corresponding
occupation probability distribution $\bar{P}_{q}=\bar{\rho}_{qq}$
extends over more than fifteen states Fig.~\ref{fig:fig4}(c)
(squares).  We then perform a numerical fit of $\bar{\rho}_{qq'}$ on a
thermal distribution in Eq.~(\ref{eq:effective}). The fit leads to an
effective temperature $\tau_{\mathrm eff}$ with an error $\Delta\tau$,
signaling the departure from the approximatively diagonal thermal
density matrix.  In considered diabatic regime, $\gamma \ll 1$,
$\tau_{\mathrm{eff}}\approx\tau$ with a very small relative error
$\delta\tau=\Delta\tau/\tau_{\mathrm{eff}}<10^{-4}$.\\
\noindent Increasing $\gamma$, the {\em coherent regime} is entered,
with a vibron occupation probability considerably altered, see
Fig.~\ref{fig:fig4}(b). We observe two main modifications.  First,
{\em off-diagonal} elements are {\em larger} than in the diabatic
regime, even though they remain rather small in comparison with the
diagonal ones. Second, the probability distribution along the diagonal
gets narrower.  Remarkably, the occupation probabilities $\bar{P}_{q}$
are still approximated by a quasi-thermal state {\em but with an
  effective temperature $\tau_{\mathrm{eff}}$ lower than the
  environmental one}, see Fig.~\ref{fig:fig4}(c) circles. The relative
error is still rather small, $\delta\tau\approx 6\cdot10^{-3}$.  We remark
that this result is a consequence of the non-vanishing vibronic
coherences, in fact in the RWA the density matrix would be strictly
diagonal, as a difference with Fig.~\ref{fig:fig4}(b).

\begin{figure}[ht]
\begin{center}
\includegraphics[width=7cm,keepaspectratio]{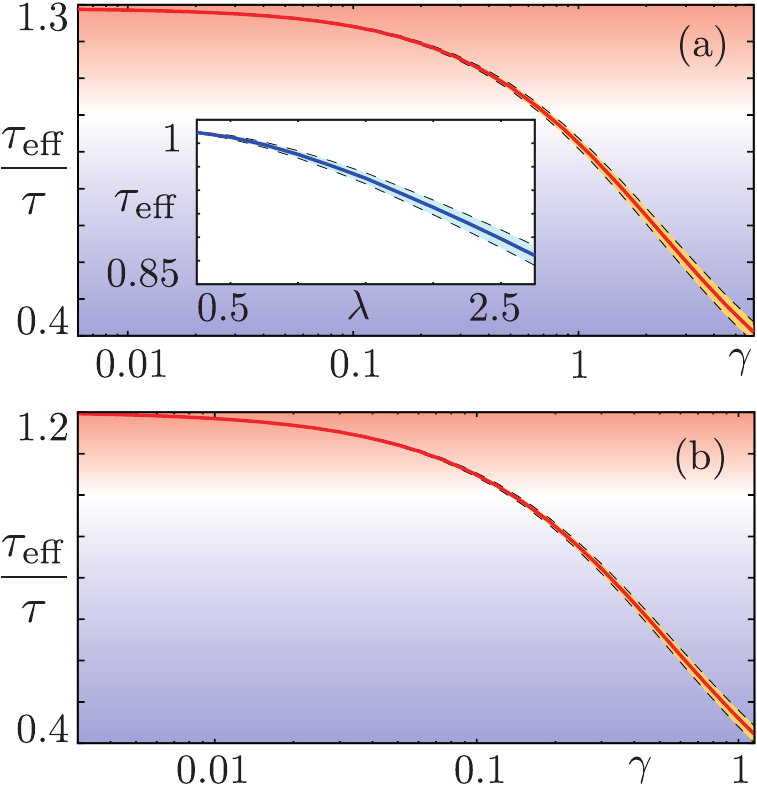}
\caption{(Color online) Solid lines: effective temperature
  $\tau_{\mathrm{eff}}$ as a function of $\gamma$, extracted from a
  numerical fit of $\bar{\rho}_{qq'}$ on Eq.~(\ref{eq:effective}),
  (see text). Panel (a): $\tau=9$ and $eV=20\,\omega_{0}$. In the
  inset $\tau_{\mathrm{eff}}$ as a function of $\lambda$ for
  $\gamma=6$.  Panel (b): $\tau=3$ and $eV=6\,\omega_{0}$. In all
  panels, dashed lines around the solid line delimit the absolute error
  on the effective temperature, $\Delta\tau$. Red (blue) gradient
  areas signal heating (cooling). Other parameters are
  $n_{\rm{g}}=1/2$ and $\omega_{\rm{c}}=10^6\ \omega_{0}$.}
\label{fig:fig5}
\end{center}
\end{figure}
\noindent 
The crossover from heating to cooling with increasing $\gamma$ and the
accuracy of the thermal approximation measured by the error of the
fitting procedure are analyzed in Fig.~\ref{fig:fig5}(a)~-~(b). The
absolute error is represented by the shaded area around the continuous
line limited by dashed lines. Two different regimes are clearly
identified. In the diabatic regime, the effective temperature is {\em
  larger} than that of the environment (red shaded region) and the
system exhibits heating. The values of $\tau_{\mathrm{eff}}\geq\tau$
are due to the considered high voltage bias and indeed for $V\to 0$
one obtains $\tau_{\mathrm{eff}}\to\tau$.\\
\noindent As $\gamma$ is increased, cooling occurs (blue shaded
region) and $\tau_{\mathrm{eff}}$ drops markedly below the electronic
temperature $\tau$. Even though $\tau_{\mathrm{eff}}$ increases for
increasing $V$ (not shown) this cooling effect survives up to
$eV\approx 20\,\omega_{0}$. It can be seen that the error grows with
increasing $\gamma$, signaling the rise of the off-diagonal terms of
the density matrix. We can then safely identify the cooling phenomenon
provided $\gamma$ does not approach $\tau$. In the inset of
Fig~\ref{fig:fig5}(a) the typical behavior of $\tau_{\mathrm{eff}}$ as
a function of $\lambda$ is shown, with an error on $\tau_{\mathrm{eff}}$ which increases increasing $\lambda$.\\
\noindent We remark that the cooling phenomenon is entirely due to the
coherent dynamics of the vibron-electronic system and is not induced
by any ad-hoc mechanism acting on the system.\\

\begin{figure}[ht]
\begin{center}
\includegraphics[width=8cm,keepaspectratio]{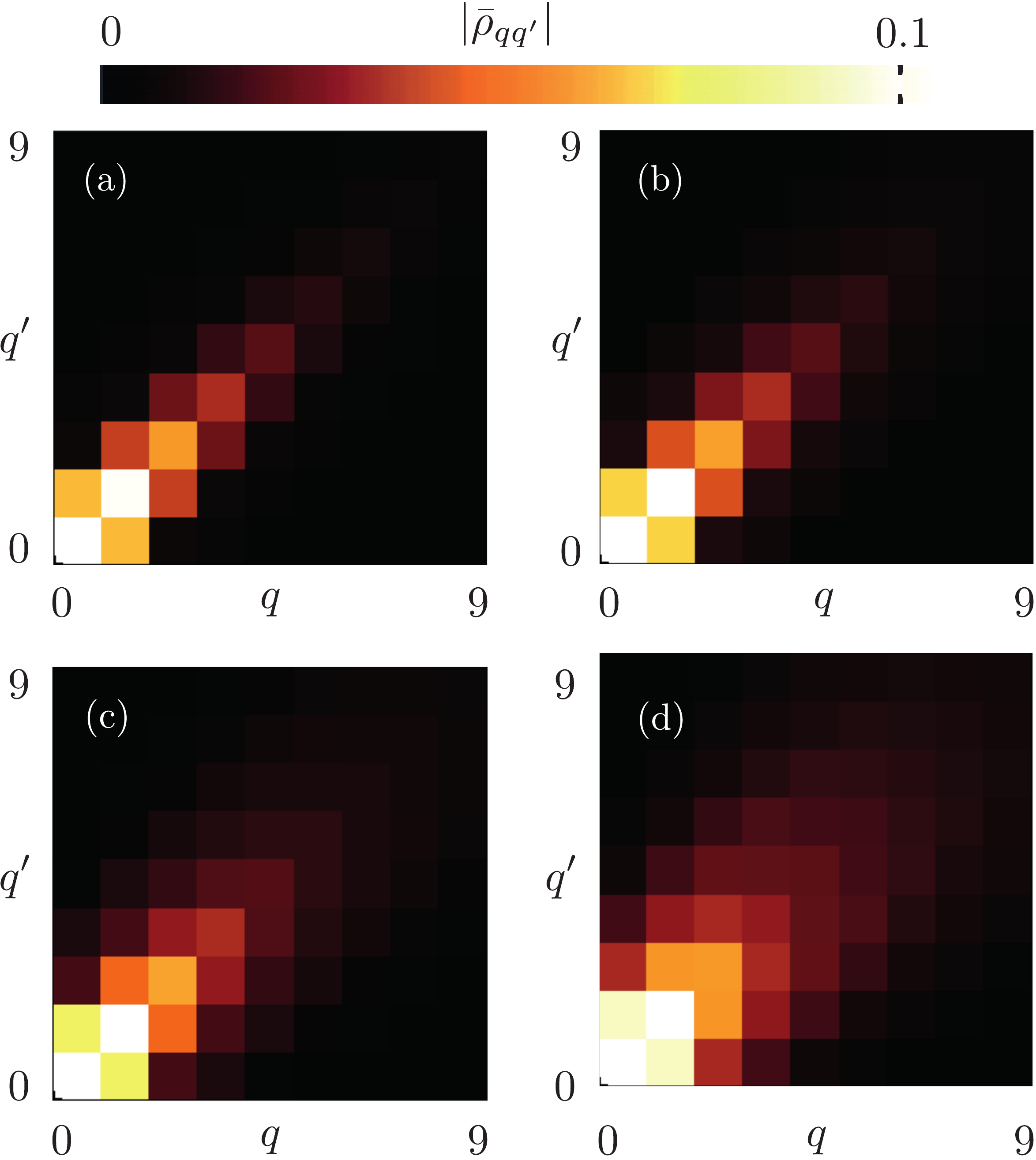}
\caption{(Color online) Density plot of $|\bar{\rho}_{qq'}|$ as a
  function of $q,q'$ for fixed $\gamma$, $\tau$ and different values
  of the electron-vibron coupling strength.  Panel (a): $\lambda=1$;
  Panel (b): $\lambda=1.5$; Panel (c): $\lambda=2$; Panel (d):
  $\lambda=2.5$. In all panels, $n_{\rm{g}}=1/2$, $eV=\omega_{0}$,
  $\gamma=3$, $\tau=6$ and $\omega_c=10^6\ \omega_{0}$.}
\label{fig:fig6}
\end{center}
\end{figure}

\noindent The description in terms of an effective thermal
distribution ceases to be valid when $\gamma\to\tau$, as shown by the
increasing errors in Fig.~\ref{fig:fig5}. To investigate this regime,
we consider the case $\tau=3$, $\gamma=1.2$ in
Fig.~\ref{fig:fig4}(d). Here, although an effective temperature
$\tau_{\mathrm{eff}}\approx 0.8<\tau$ can be formally extracted with
the fitting procedure, the relative error becomes large
$\delta\tau\approx0.11$.  The main source of error are the rather
large off-diagonal matrix elements. This is a general trend which we
always observed when $\gamma$ approaches $\tau$. As an illustration,
we report in the inset of Fig.~\ref{fig:fig4}(c) the density matrix
for the same parameters of the main panel but at higher temperature
($\tau\gg\gamma$). In this case off-diagonal elements are strongly
suppressed and an effective thermal description is then appropriate.\\
\noindent We conclude this paragraph commenting on the dependence of
the above results on the electron-vibron coupling strength, $\lambda$,
which is not limited in our approach.  In Figure~\ref{fig:fig6}
(a)~-~(d) we report the density plot of $|\bar{\rho}_{qq'}|$ at fixed
$\tau$ and for increasing values of $\lambda$. Increasing the
dot-vibron coupling induces an hybridization of the electronic and
mechanical degrees of freedom which manifests itself also in the
off-diagonal elements of the vibron density matrix between Fock
states.  Indeed, with increasing $\lambda$ a ``delocalization"
phenomenon is induced, the density matrix spreading away from the
diagonal.

\subsection{Wigner function and quadratures}
Further insight on the system behavior is obtained from the Wigner
quasi-probability distribution function~\cite{mandel}
\begin{equation}
W(x,p)=\frac{1}{\pi}\int_{-\infty}^{\infty}{\mathrm
  d}y\ e^{2ipy}\langle
x-y|\mathcal{U}^{\dagger}\bar{\rho}\mathcal{U}|x+y\rangle\, ,
\end{equation}
the quantum analogue of the Liouville density in classical phase space
with $\mathcal{U}$ defined in Eq.~(\ref{eq:platrafo}). The Wigner
function allows the detection of non-classical features signaled by
regions in the $(x,p)$ plane where $W(x,p)<0$.~\cite{mandel}
\begin{figure}[ht]
\begin{center}
\includegraphics[width=8cm,keepaspectratio]{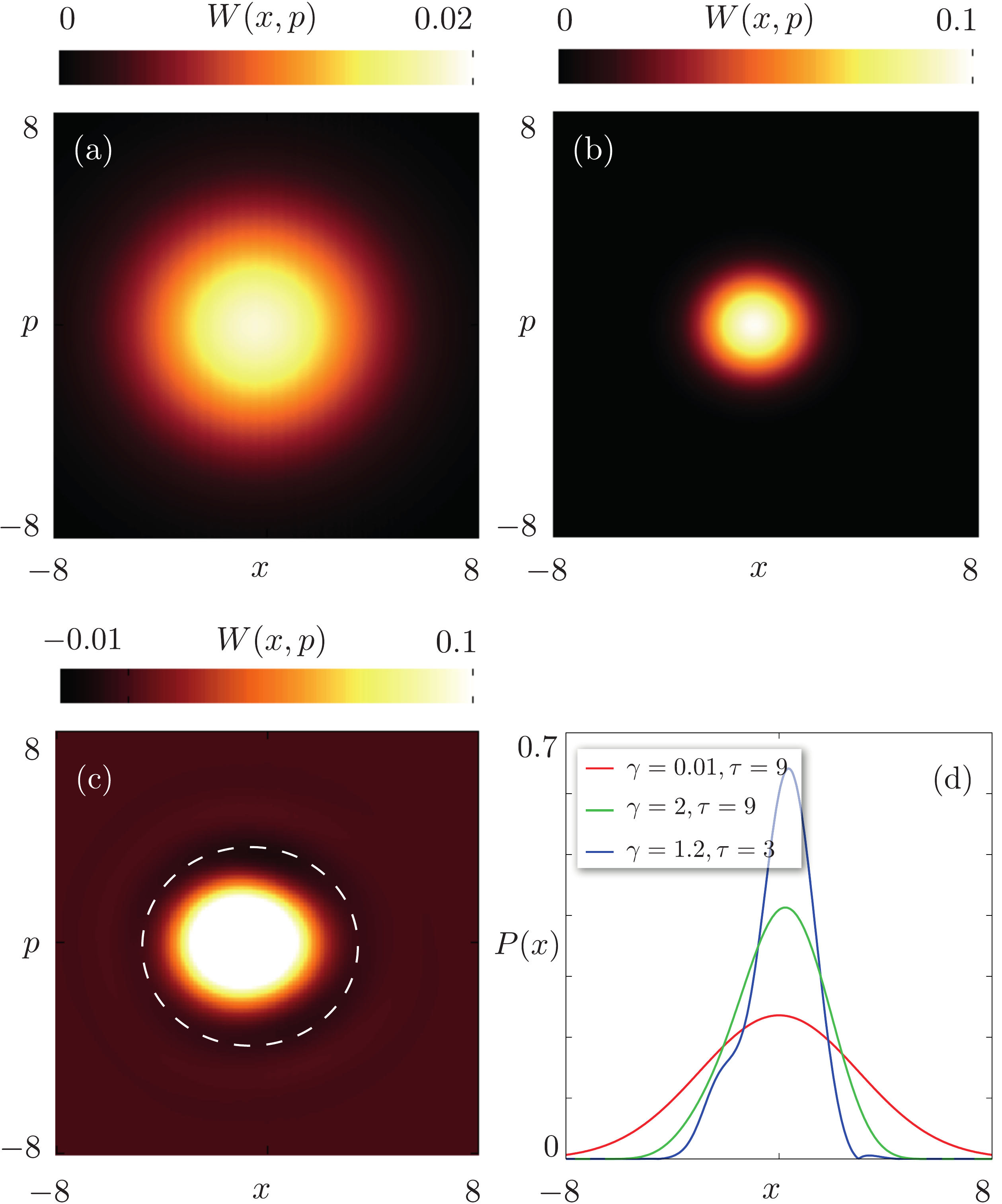}
\caption{(Color online) Panels (a)~-~(c) density plot of the Wigner
  function $W(x,p)$ as a function of $x,p$.  In panels (a) and (b)
  $\tau=9$, and in (a) $\gamma=0.01$, in (b) $\gamma=2$.  Note the
  different color scale in the two panels. (c) Density plot of
  $W(x,p)$ for $\tau=3$ and $\gamma=1.15$. (d) Probability density of
  the oscillator position $P(x)$. In all panels $x$ and $p$ are
  expressed in unity of $\ell_0$ and $\ell_0^{-1}$ respectively - see
  Eq.~(\ref{eq:ell0}). In addition $n_{\rm{g}}=1/2$, $eV=\omega_{0}$,
  $\lambda=2$ and $\omega_c=10^6\ \omega_{0}$.}
\label{fig:fig7}
\end{center}
\end{figure}
\noindent

Figures~\ref{fig:fig7}(a)~-~(b) show $W(x,p)$ for the parameters of
Figs.~\ref{fig:fig4}(a)~-~(b), in the effective thermal regime. The
Wigner functions are positive and can be regarded as the probability
density of the oscillator states in the $(x,p)$ phase
space. Consistently, their shape closely resembles that of a thermal
state,~\cite{mandel} but with a decreased width
$\tau_{\mathrm{eff}}$. The shrinkage of $W(x,p)$ observed when
entering the coherent regime can be traced back to the role of
coherences which mediate the redistribution of the $\bar{P}_{q}$ and
the ensuing reduction of $\tau_{\mathrm{eff}}$.\\
\noindent Figure~\ref{fig:fig7}(c) shows $W(x,p)$ for $\gamma=1.2$ and
$\tau=3$ in the {\em strongly coherent} regime, when the system can no
longer be described by a thermal distribution (as in
Fig.~\ref{fig:fig4}(d)). In this case, the width of $W(x,p)$ is still
reduced. As a difference with previous case, here we observe that
$W(x,p)$ becomes negative in an approximately circular area around the
main peak (white dashed circle). This fact signals a {\em
  non-classical} behavior of the system entirely due to the
non-diagonal structure of the reduced density matrix. Negative values
of $W(x,p)$ are a trademark of the regime $\gamma/\tau\approx 1$.
This quantum behavior, naively unexpected at the considered high
temperatures $\tau\gtrsim 1$, is a relevant result whose implications
on the vibronic fluctuations will be discussed in the next
subsection.\\
\noindent Also noteworthy is the elongated shape of $W(x,p)$ along the
$x$ direction, Fig.~\ref{fig:fig7}(c). This is reflected in the
position probability distribution
\begin{equation}
P(x)=\int_{-\infty}^{\infty}{\mathrm d}p\ W(x,p)
\end{equation}
shown in Figure~\ref{fig:fig7}(d). While in the effective-thermal
regime $P(x)$ exhibits a single peak, in the regime of strong
coherences it shows a peculiar shoulder structure. In this latter
regime, for relatively slow tunneling events $\gamma \sim 1$, the
oscillator adjusts itself to the two equilibrium positions
corresponding to zero or one extra electron on the dot, $x_{0}=\lambda
n_{\mathrm g}$ and $x_{1}=\lambda(n_{g}-1)$ (in units of $\ell_0$, see
Eq.~(\ref{eq:ell0})). When the width of the two probability
distributions for the uncharged and charged states ($\propto
\tau_{\mathrm{eff}}$) is of the order of the separation between the
rest positions $\propto \lambda$, the shoulder is visible. The effect
is therefore more pronounced the larger is the electron-vibron
coupling $\lambda$ and eventually develops into a double-peaked shape.

\begin{figure}[ht]
\begin{center}
\includegraphics[width=7cm,keepaspectratio]{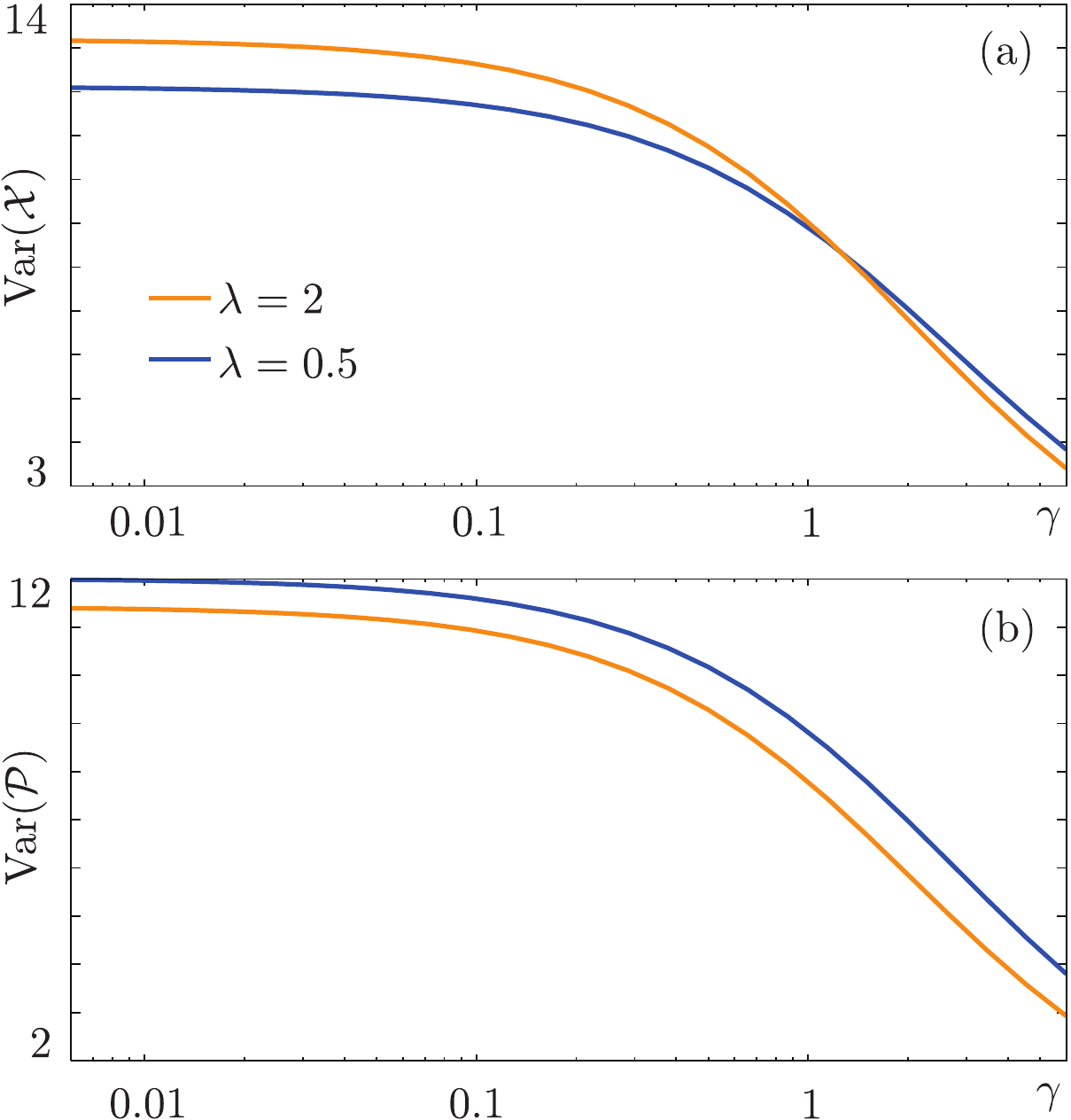}
\caption{(Color online) Variance of the position $\mathcal{X}$ (a) and
  of the momentum $\mathcal{P}$ (b) as a function of $\gamma$.  In all
  panels $\tau=9$, $eV=20\,\omega_{0}$, $n_{\mathrm{g}}=1/2$ and
  $\omega_{c}=10^{6}\ \omega_{0}$.}
\label{fig:fig10}
\end{center}
\end{figure}

\noindent The width of $W(x,p)$ in the $x$ and the $p$ directions is
strictly related to the variance of the vibron position and momentum
\begin{equation}
\langle x^{\mu}\rangle=\int_{-\infty}^{\infty}{\mathrm d}x\int_{-\infty}^{\infty}{\mathrm d}p\ x^{\mu}W(x,p)\, ,
\end{equation}
and corresponding expressions for $p$. In Fig~\ref{fig:fig10} (a) and
(b) are reported the variance of the position,
$\mathcal{X}=x/\ell_0=(b^{\dag}+b)/\sqrt{2}$, and momentum
$\mathcal{P}=p\cdot \ell_0=i(b^{\dagger}-b)/\sqrt{2}$ satisfying
always $\mathrm{Var}(\mathcal{X})\mathrm{Var}(\mathcal{P})\geq1/2$ -
with $\ell_{0}$ defined in Eq.~(\ref{eq:ell0}).
As expected, in the coherent regime both variances are strongly
suppressed with respect to the large values attained in the diabatic
regime, both for high and low (not shown) voltages. This fact denotes
a \emph{tendency} towards {\em squeezing}.~\cite{mandel}

\subsection{Vibronic Fluctuations}
\noindent 
The out-of-equilibrium vibron behavior is conveniently discussed
introducing the {\em vibron} Fano factor~\cite{merlo,rod}
\begin{equation}
F_{\rm{v}} =\frac{\mathrm{Var}(N)}{\langle N\rangle}\, ,
\label{eq:fano}
\end{equation}
where $N \equiv b^{\dag}b$. Here, $F_{\rm{v}}$ is the ratio between
the variance of the vibron occupation number $\mathrm{Var}(N)=\langle
N^2\rangle-\langle N \rangle^2$ and its average $\langle N\rangle$,
where $\langle\mathcal{O}\rangle=\mathrm{Tr}\{\mathcal{O}\rho\}
=\mathrm{Tr}\{\mathcal{U}\mathcal{O}\mathcal{U}^\dagger\bar{\rho}\}$.
This quantity (not be confused with the \emph{electronic} Fano factor)
brings information about the statistics of the vibronic mode and also
on electronic properties, like charge fluctuations.  This is due to
the involved polaron transformation and it can be explicitly seen
introducing the hybrid average
$\langle\mathcal{O}\rangle_{\bar\rho}=\rm{Tr}\{\mathcal{O}\bar{\rho}\}$.
In terms of $\langle\cdot\rangle_{\bar{\rho}}$ one has
\begin{eqnarray}
\begin{split}
\label{eq:avg}
\langle N \rangle=&\,\langle N \rangle_{\bar{\rho}}-\sqrt{2}\langle\hat{\eta}\mathcal{X}\rangle+\langle\hat{\eta}^2\rangle\\
\mathrm{Var}(N)=&\,\mathcal{V}(N)+\mathcal{V}(\hat{\eta}^2)+2\mathcal{V}(\hat{\eta}\mathcal{X})+2\mathcal{C}(\hat{\eta}^2,N)\\
&-\sqrt{2}\left[\mathcal{C}(\hat{\eta}\mathcal{X},N)+\mathcal{C}(N,\hat{\eta}\mathcal{X})+2\mathcal{C}(\hat{\eta}^2,\hat{\eta}\mathcal{X})\right]\,,
\end{split}
\end {eqnarray}
where
$\mathcal{V}(\mathcal{O})\equiv\langle\mathcal{O}^2\rangle_{\bar{\rho}}-\langle\mathcal{O}\rangle^2_{\bar{\rho}}$
and
$\mathcal{C}(\mathcal{A},\mathcal{B})\equiv\langle\mathcal{A}\mathcal{B}\rangle_{\bar{\rho}}-\langle\mathcal{A}\rangle_{\bar{\rho}}\langle\mathcal{B}\rangle_{\bar{\rho}}$,
with adimensional oscillator position $\mathcal{X}$. The quantities
$\langle N \rangle$ and $\mathrm{Var}(N)$ depend explicitly both on
the charge (being $\hat{\eta}=\lambda(\hat{n}-n_{\rm{g}})$) and on its
fluctuations.  In Eq.~(\ref{eq:avg}) all terms of the form
$\langle\cdot\,\mathcal{X}\rangle$ and
$\mathcal{C}(\cdot,\,\cdot\,\mathcal{X})$ depend on the coherences of
the density matrix.  In the RWA these terms simplify but keep the
$\hat{\eta}$ dependence.  We remark that, beside the explicit
dependence of $\mathrm{Var}(N)$ and $\langle N\rangle$ on $\langle
n\rangle$, fluctuations of the electronic charge intrinsically affect
$F_{\rm{v}}$, since the contribution of electronic and vibronic
degrees of freedom cannot be factorized in $\bar{\rho}_{qq'}^n$.\\
\noindent $F_{\rm{v}}$ belongs to the class of ``bosonic" Fano factors,
originally introduced to characterize \emph{boson} distributions and
often used to study photon populations in the context of quantum
optics.~\cite{mandel} Typically, for bosons the Fano factor is
super-Poissonian ($F_{\rm{v}}>1$). This is, for example, the situation
for light from a classical radiation field.~\cite{mandel} Deviations
from the super-Poissonian regime signal non-classical correlations,
which have been observed for instance in experiments with
micromaser.~\cite{maser} From an experimental point of view, the
situation for vibron is more complicated than in the case of photons,
although many proposals have been made to observe quantities like the
average vibron number.~\cite{santa,buks,woolley} Recent experiments
aiming at characterizing the vibron distributions have
appeared.~\cite{o_connel} A few recent theoretical works investigated
the vibron distributions predicting sub-Poissonian vibron Fano factor
{\em in the diabatic regime} $\gamma\ll 1$, for low temperatures
$\tau\ll 1$ and in selected parameter ranges.~\cite{merlo,cava3}\\
\begin{figure}[ht]
\begin{center}
\includegraphics[width=7cm,keepaspectratio]{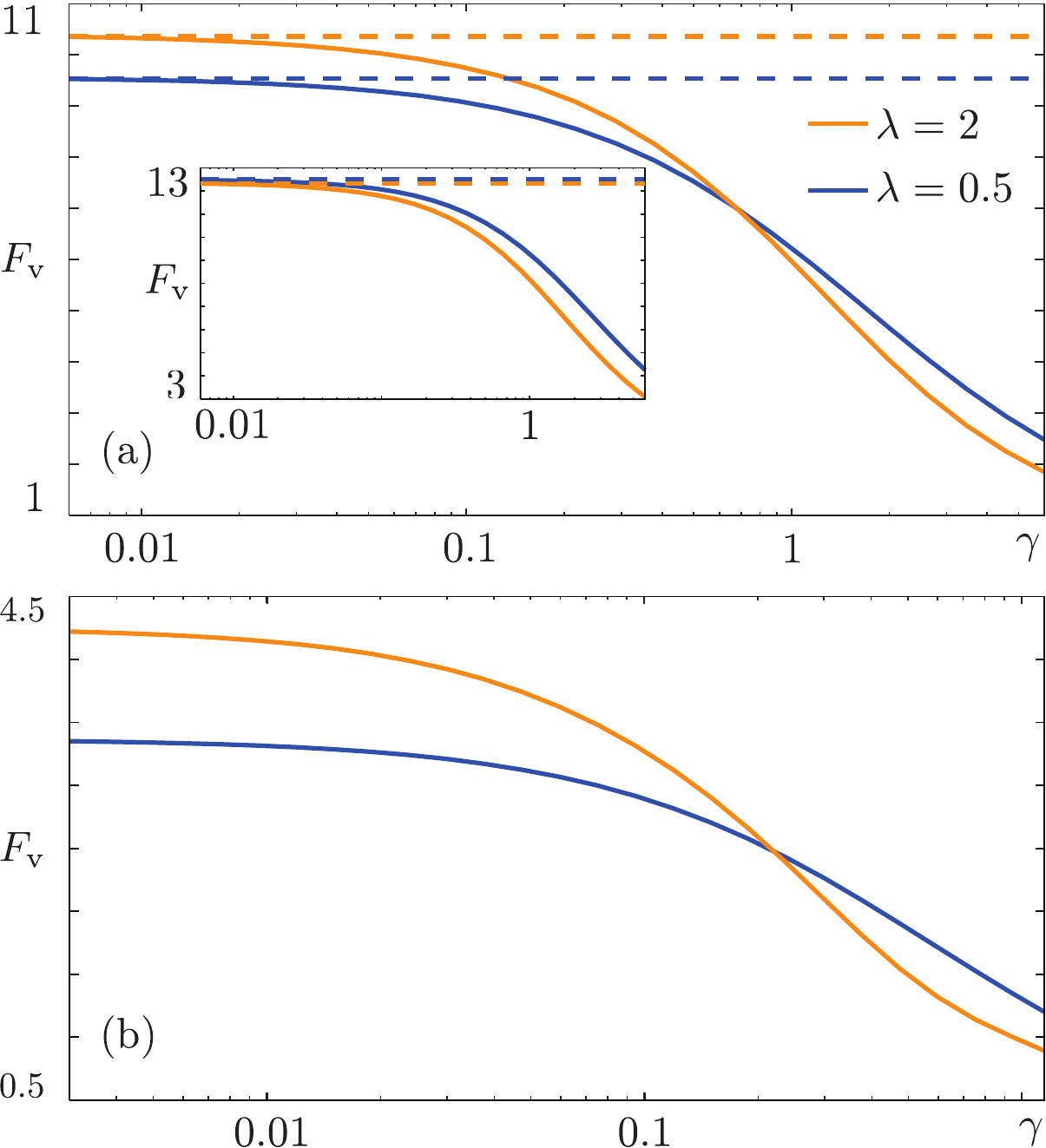}
\caption{(Color online) Vibron Fano factor $F_{\rm{v}}$ as a function
  of $\gamma$ for different values of the
  electron-vibron coupling strength $\lambda$ and temperature
  $\tau$. In panel (a) $\tau=9$ and $eV=\omega_{0}$, in the inset
  $eV=20\ \omega_{0}$.  In panel (b) $\tau=3$ and $eV=\omega_{0}$.
  Solid lines are the results obtained with the GME, dashed lines are
  obtained within the RWA. In all panels, $n_{\rm{g}}=1/2$ and
  $\omega_{\rm{c}}=10^6\ \omega_{0}$.}
\label{fig:fig8}
\end{center}
\end{figure}
\noindent The behavior of the vibron Fano factor as a function of
$\gamma$ for weak and strong $\lambda$ is reported in
Fig.~\ref{fig:fig8}. In the regimes where the vibron can be
approximated by an effective thermal distribution, $F_{\rm{v}}$
displays a super-Poissonian behavior. This occurs in the diabatic
regime, both at low and high voltages ($e V$ smaller or larger than
$k_{\rm{B}} T$), and in the weakly coherent regime. In this case however
the vibron Fano factor is considerably reduced. These behaviors
qualitatively do not depend on the electron-vibron coupling
$\lambda$. In the diabatic regime the RWA applies with no dependence
on $\gamma$.\\
\noindent On the other hand, entering the strongly coherent regime
($\gamma\approx 1$ and $\tau=3$ in Fig.~\ref{fig:fig8}(b)),
non-classical sub-Poissonian vibron Fano factors are obtained.  This
is in correspondence with the negative values of the Wigner function
observed in this parameter regime.  Both features result from the
large vibronic coherences.  Similar conclusions were drawn from the
frequency-dependent shot noise of a NEMS.~\cite{brandes1} Here,
however, these peculiar quantum behaviors show up in the steady state
regime.\\
\begin{figure}[ht]
\begin{center}
\includegraphics[width=7cm,keepaspectratio]{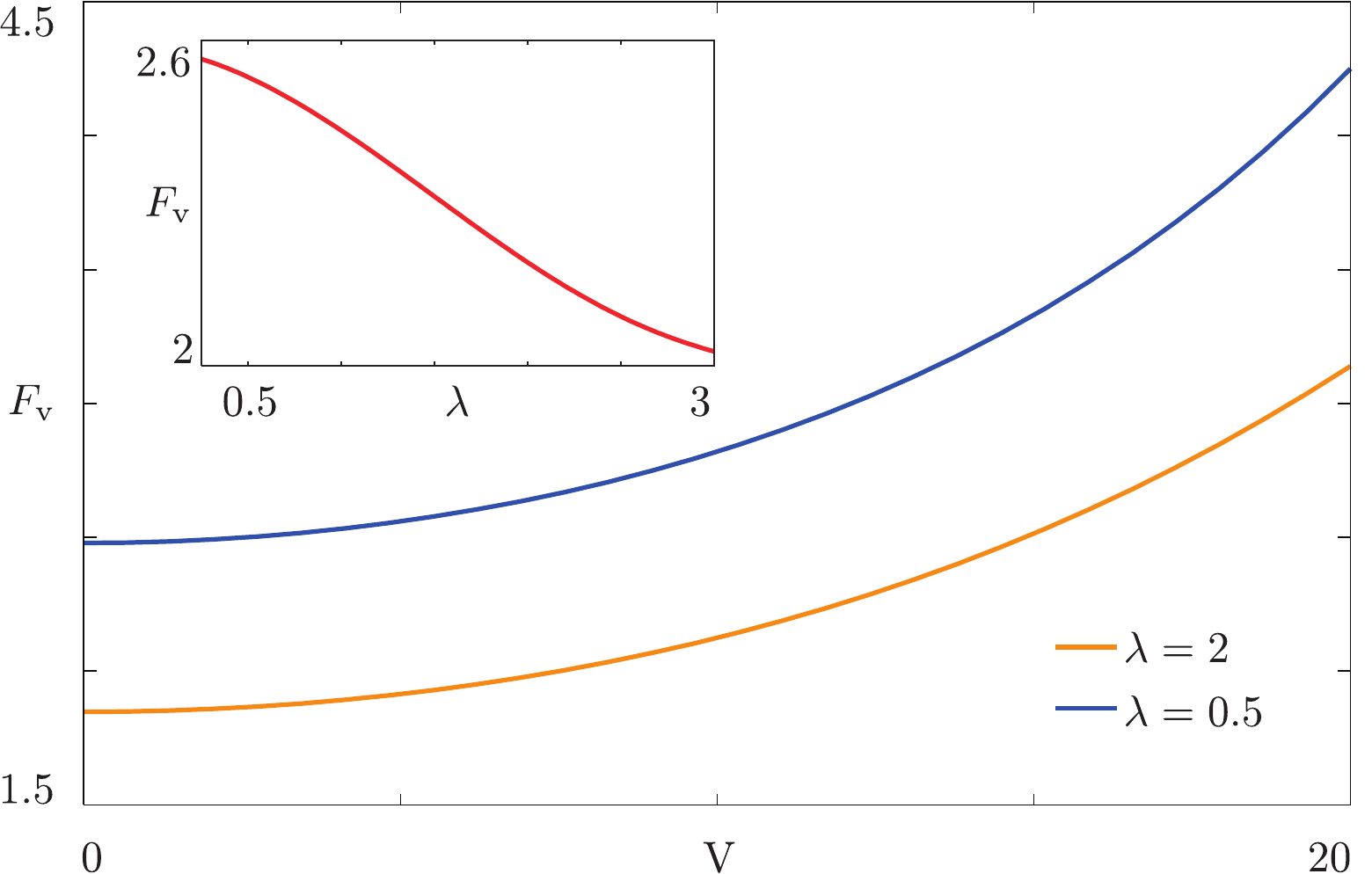}
\caption{(Color online) Main: plot of $F_{\rm{v}}$ as a function of $V$ (units
  $\omega_{0}/e$) for different values of $\lambda$. Inset: dependence
  of $F_{\rm{v}}$ on the electron-vibron strength $\lambda$, for
  $eV=15\,\omega_{0}$. Other parameters are $\gamma=6$, $n_{\rm{g}}=1/2$,
  $\tau=9$ and $\omega_{\rm{c}}=10^{6}\ \omega_{0}$.}
\label{fig:fig9}
\end{center}
\end{figure}
\noindent 
The vibron Fano factor also depends on the voltage and on
electron-vibron coupling.  $F_{\rm{v}}$ increases with increasing bias
voltage, assuming super-Poissonian values even in the coherent regime,
Fig.~\ref{fig:fig9} (main panel).  This behavior can be attributed to
the increase of scattering between oscillator states induced by the
current flow across the system.  In the inset, the Fano factor is
plotted as a function of $\lambda$: $F_{\rm{v}}$ {\em decreases} with
increasing coupling strengths.  This trend has been found both in the
low ($eV<k_{\rm B}T$) and in the high ($eV>k_{\rm B}T$) voltage
regimes - see also Figs.~\ref{fig:fig8}(a)~-~(b).\\

\subsection{Charge degree of freedom}
\begin{figure}[ht]
\begin{center}
\includegraphics[width=7.5cm,keepaspectratio]{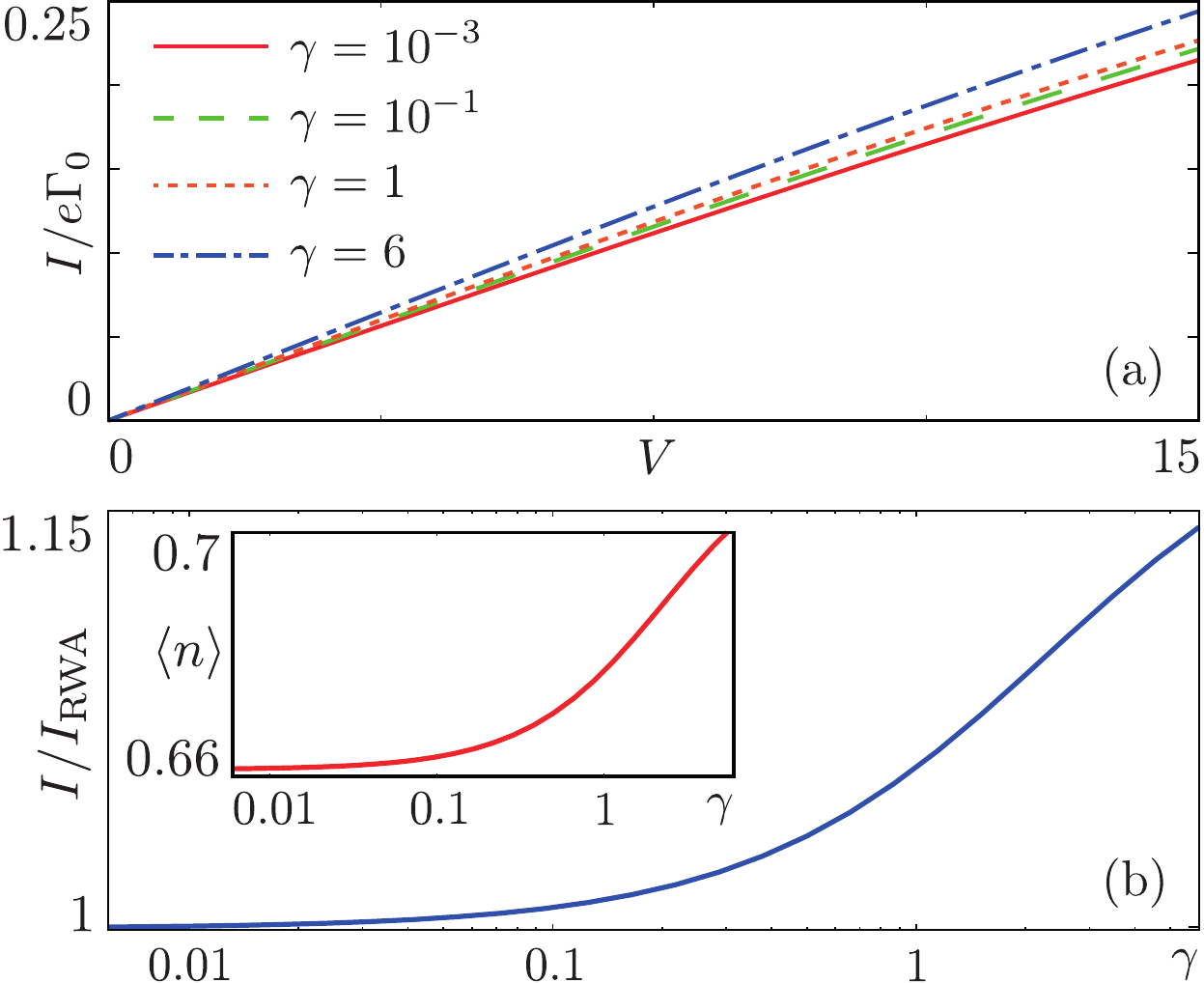}
\caption{(Color online) (a) Current $I$ (units $e\Gamma_{0}$) as a
  function of $V$ (units $\omega_{0}/e$) for different values of
  $\gamma$. (b) Ratio between the current obtained with the GME ($I$)
  and the current in the RWA approximation ($I_{\rm{RWA}}$) as a
  function of $\gamma$, at $eV=15\,\omega_{0}$. Inset: average
  occupation $\langle n\rangle$ of the electronic level. Other
  parameters: $n_{\rm{g}}=1/2$, $\tau=9$, $\lambda=2$ and
  $\omega_c=10^6\ \omega_{0}$.}
\label{fig:fig11}
\end{center}
\end{figure}

We conclude the survey of our results briefly commenting on the
electronic properties. Figure~\ref{fig:fig11}(a) shows the $I(V)$
characteristics for different values of $\gamma$. In the regime
$\tau>1$, the vibron sideband can not be resolved and the current
increases monotonically.
\noindent Figure~\ref{fig:fig11}(b) shows the ratio between the
current $I$ obtained from the GME and the current in the RWA,
$I_{\rm{RWA}}$, as a function of $\gamma$ and fixed voltage
$V$. Increasing $\gamma$, the current increases with respect to the
RWA limit. This qualitative trend occurs in all the parameter regimes
explored.  The increase of the current is more prominent for larger
values of $\lambda$. This is a consequence of the delocalization of
the vibron density matrix occurring with increasing electron-vibron
coupling.\\
\noindent The average occupation of the electronic dot level $\langle
\hat n\rangle$ behaves similarly to the current (Fig.~\ref{fig:fig11}
inset).  For the case of only $n=0$ and $n=1$ electron states the
variance of the electron occupation number reads
\begin{equation}
\mathrm{Var}(\hat n)=\langle \hat n^{2}\rangle-\langle \hat n\rangle^2=\langle
\hat n\rangle\left(1-\langle \hat n\rangle\right)\, .
\end{equation}
The increase of $\langle \hat n\rangle>1/2$ signals a
suppression of the fluctuations of the electronic level population
analogously to the fluctuations of the vibronic part.\\

\section{conclusions}
\label{sec:concl}
In the present article we derived a generalized master equation to
explore the steady-state properties of a nano-electromechanical system
in a wide parameters range.  
extends from the very fast vibrations $\omega_{0}\gg\Gamma_{0}$ to the
slow, coherent regime where the {\em off-diagonal} elements of the
reduced density matrix between energy eigenstates are paramount.\\
\noindent In the coherent regime, two peculiar behaviors have been
found. For intermediate frequencies, $\omega_{0}\approx\Gamma_{0}$,
the system can be described in terms of an {\em effective} thermal
distribution with a temperature lower than that of the environment.
The cooling phenomenon is accompanied by a decrease of position and
momentum quadratures.  For still slower oscillations, a strongly
coherent regime is entered characterized by non classical behavior.  A
benchmark of this regime is a marked suppression of the vibron Fano
factor, which can even attain {\em sub-Poissonian} values. \\
\noindent This work is one of the first steps towards the
understanding of dynamical properties of nano-electromechanical systems
in the coherent regime and represents a rather tough numerical
challenge, due to the slow convergence of the master equation solution
and the need of a large number of basis states. Future investigations
are certainly in order, to explore the regime of very strong
electron-vibron coupling and the crossover towards the semi-classical
regime. Further interesting issues to be investigated are higher order 
electronic properties, like the current fluctuations and
their connection with the fluctuations of the mechanical part.\\

\noindent\textit{Acknowledgments.} The authors acknowledge stimulating
discussions with A. Nocera. Financial support by CNR-SPIN via both the
Seed Project PLASE001 and the ``Progetto giovani'', and by the EU-FP7
via ITN-2008-234970 NANOCTM is also gratefully acknowledged.


\begin{thebibliography}{99}
\bibitem{roukes} M. L. Roukes, \textit{Nanoelectromechanical Systems}, in \textit{Technical Digest of the 2000 Solid State Sensor and Actuator Workshop}, Transducers Research Foundation, Cleveland, OH (2000).
\bibitem{park} H. Park, J. Park, A. K. L. Lim, E. H. Anderson, A. P. Allvisatos, and P. McEuen, Nature \textbf{407}, 57 (2000).
\bibitem{let} R. Leturcq, C. Stampfer, K. Inderbitzin, L. Durrer, C. Hierold, E. Mariani, M. G. Schultz, F. von Oppen, and K. Ensslin, Nature Phys. {\bf 5}, 327 (2009).
\bibitem{leroy} B. J. Leroy, S. G. Lemay, J. Kong, and C. Dekker, Nature {\bf 432}, 371 (2004).
\bibitem{sap} S. Sapmaz, P. Jarillo-Herrero, Y. M. Blanter, C. Dekker, and H. S. J. van der Zant, Phys. Rev. Lett. {\bf 96}, 026801 (2006).
\bibitem{lass} B. Lassagne, Y. Tarakanov, J. Kiranet, D. Garcia-Sanchez, and A. Bachtold, Science \textbf{325}, 1107 (2009).
\bibitem{knobel} R. G. Knobel and A. N. Cleland, Nature {\bf 424}, 291 (2003).
\bibitem{nanobeams} E. M. Weig, R. H. Blick, T. Brandes, J. Kirschbaum, W. Wegscheider, M. Bichler, and J. P. Kotthaus, Phys. Rev. Lett. \textbf{92}, 046804 (2004).
\bibitem{koch} J. Koch, F. von Oppen, and A. V. Andreev, Phys. Rev. B {\bf 74}, 205438 (2006).
\bibitem{mitra} A. Mitra, I. Aleiner, and A. J. Millis, Phys. Rev B {\bf 69}, 245302 (2004).
\bibitem{flens} S. Braig and K. Flensberg, Phys. Rev. B \textbf{68}, 205324 (2003).
\bibitem{shen} X. Y. Shen, B. Dong, X. L. Lei, and N. J. M. Horing, Phys. Rev. B \textbf{76}, 115308 (2007).
\bibitem{zazunov} A. Zazunov, D. Feinberg, and T. Martin, Phys. Rev. B {\bf 73}, 115405 (2006).
\bibitem{cava} F. Cavaliere, A. Braggio, J. T. Stockburger, M. Sassetti, and B. Kramer, Phys. Rev. Lett. {\bf 93}, 036803 (2004).
\bibitem{braggio} A. Braggio, M. Sassetti, and B. Kramer, Phys. Rev. Lett. {\bf 87}, 146802 (2001).
\bibitem{cava1} F. Cavaliere, A. Braggio, M. Sassetti, and B. Kramer, Phys. Rev. B {\bf 70}, 125323 (2004).
\bibitem{haupt} F. Haupt, F. Cavaliere, R. Fazio, and M. Sassetti Phys. Rev. B {\bf 74}, 205328 (2006).
\bibitem{sid} J. A. Sidles, J. L. Garbini, K. J. Bruland, D. Rugar, O. Z\"uger, S. Hoen, and C. S. Yannoni, Rev. Mod. Phys. \textbf {67}, 249 (1995).
\bibitem{il} B. Ilic, H. G. Craighead, S. Krylov, W. Senaratne, C. Ober, and P. Neuzil,  J. Appl.
Phys. \textbf{95}, 3694 (2004).
\bibitem{nanoscale} A. K. Naik, M. S. Hanay, W. K. Hiebert, X. L. Feng, and M. L. Roukes, Nature Nanotech. \textbf{4}, 445 (2009). 
\bibitem{giazotto} F. Giazotto, T. Heikkil\"a,
  A. Luukanen, A. M. Savin, and J. P. Pekola, Rev. Mod. Phys. {\bf
    78}, 217 (2006).
\bibitem{mandel} L. Mandel and E. Wolf, \textit{Optical Coherence and Quantum Optics}, N. Y. : Cambridge University Press, 1995.
\bibitem{hertz} J. B. Hertzberg, T. Rocheleau, T. Ndukum, M. Savva, A. A. Clerk, and K. C. Schwab, Nature Phys. \textbf{6}, 213 (2010).
\bibitem{koch2} J. Koch and F. von Oppen, Phys. Rev. Lett. \textbf{94}, 206804 (2005).
\bibitem{mozy} D. Mozyrsky, M. B. Hastings, and I. Martin, Phys. Rev. B {\bf 73}, 035104 (2006).
\bibitem{nocera} A. Nocera, C. A. Perroni, V. Marigliano Ramaglia, and V. Cataudella, Phys. Rev. B \textbf{83}, 115420 (2011).
\bibitem{huss} R. Hussein, A. Metelmann, P. Zedler, and T. Brandes, Phys. Rev. B \textbf{82}, 165406 (2010).
\bibitem{merlo} M. Merlo, F. Haupt, F. Cavaliere, and M. Sassetti, New. J. Phys. \textbf{10}, 023008 (2008).
\bibitem{cava3} F. Cavaliere, G. Piovano, E. Paladino, and M. Sassetti, New J. Phys \textbf{10}, 115004 (2008). 
\bibitem{armour} A. D. Armour, M. P. Blencowe, and Y. Zhang, Phys. Rev. B {\bf 69}, 125313 (2004).
\bibitem{armour2} A. D. Armour, Phys. Rev. B {\bf 70}, 165315 (2004).
\bibitem{doiron} C. B. Doiron, W. Belzig, and C. Bruder, Phys. Rev. B \textbf{74}, 205336 (2006). 
\bibitem{rod2} T. J. Harvey, D. A. Rodrigues, and A. D. Armour, Phys. Rev. B {\bf 81}, 104514 (2010).
\bibitem{harvey} T. J. Harvey, D. A. Rodrigues, and A. D. Armour, 
Phys. Rev. B \textbf{78}, 024513 (2008).
\bibitem{rodr2} D. A. Rodrigues, J. imbers, T. J. Harvey, and A. D. Armour, New J. Phys. \textbf{9}, 84 (2007).
\bibitem{clerk} A. A. Clerk and S. Bennett, New J. Phys. \textbf{7}, 238 (2005). 
\bibitem{kou} L. P. Kouwenhoven, C. M. Marcus, P. L. McEuen, S. Tarucha, R. M. Westervelt, 
N. S. Wingreen \textit{Electron transport in quantum dots}, 
edited by L. L. Sohn, L. P. Kouwenhoven, and G. Sch\"on, Kluwer Series E345, 1997.
\bibitem{ouy} S. H. Ouyang, J. Q. You, and F. Nori, Phys. Rev. B \textbf{79}, 075304 (2009).
\bibitem{rodr} D. A. Rodrigues and A. D. Armour, New J. Phys. \textbf{7}, 251 (2005).
\bibitem{pisto_c} F. Pistolesi, J. of Low Temp. Phys. \textbf{154}, 199 (2009).
\bibitem{pisto} F. Pistolesi, Ya. M. Blanter, and I. Martin, Phys. Rev. B {\bf 78}, 085127 (2008).
\bibitem{schultz} M. G. Schultz, Phys. Rev. B \textbf{82}, 195322 (2010).
\bibitem{izu} W. Izumida and M. Grifoni, New J. Phys. \textbf{7}, 244 (2005).
\bibitem{flens2} K. Flensberg, New J. Phys \textbf{8}, 5 (2006).
\bibitem{cava2} F. Cavaliere, E. Mariani, R. Leturcq, C. Stampfer, and M. Sassetti, Phys. Rev. B {\bf 81}, 201303(R) (2010).
\bibitem{galp3} M. Galperin, A. Nitzan, and M. A. Ratner, Phys. Rev. B {\bf 73} 045314 (2006).
\bibitem{galp4} M. Galperin, M. A. Ratner, and A. Nitzan, Nano Lett. {\bf 5} 125 (2005).
\bibitem{rae} I. Wilson-Rae, N. Nooshi, W. Zwerger, and T. J. Kippenberg, Phys. Rev. Lett. \textbf{99}, 093901 (2007).
\bibitem{teufel} J. D. Teufel, J. W. Harlow, C. A. Regal, and K. W. Lehnert, Phys. Rev. Lett. \textbf{101}, 197203 (2008).
\bibitem{teufel2} J. D. Teufel, T. Donner, M. A. Castellanos-Beltran, J. W. Harlow, and K. W. Lehnert, Nature Nanotech. \textbf{4}, 820 (2009).
\bibitem{rabl} P. Rabl, Phys. Rev. B \textbf{82}, 165320 (2010).
\bibitem{zippilli} S. Zippilli, A. Bachtold, and G. Morigi, Phys. Rev. B \textbf{81},  205408 (2010).
\bibitem{ruskov} R. Ruskov, K. Schwab, and A. N. Korotkov, Phys. Rev. B \textbf{71}, 235407 (2005).
\bibitem{clerk2} A. A. Clerk, F. Marquardt, and K. Jacobs, New J. Phys. \textbf{10}, 095010 (2008).
\bibitem{boubsi} R. El Boubsi, O. Usmani, and Y. M. Blanter, New J. Phys. \textbf{10}, 095011 (2008).
\bibitem{brandesfano} R. Sanchez, G. Platero, and T. Brandes, Phys. Rev. Lett {\bf 98}, 146805 (2007).
\bibitem{brandesfano2} R. Sanchez, G. Platero, and T. Brandes, Phys. Rev. B {\bf 78}, 125308 (2008).
\bibitem{usmani} O. Usmani, Y. M. Blanter, and Y. V. Nazarov, Phys. Rev. B {\bf 75}, 195312 (2007).
\bibitem{mozy2} D. Mozyrsky, I. Martin, and M. B. Hastings, Phys. Rev. Lett. {\bf 92}, 018303 (2004).
\bibitem{galp5} M. Galperin, A. Nitzan, and M. A. Ratner, J. Phys.: Condens. Matter {\bf 20}, 374107 (2008).
\bibitem{blum}K. Blum, {\it Density Matrix Theory and Application}, Plenum Press, New York (1996).
\bibitem{brandes1} H. Huebener and T. Brandes, Phys. Rev. Lett. \textbf{99}, 247206 (2007).
\bibitem{rod} D.A. Rodrigues, J. Imbers, and A. D. Armour, Phys. Rev. Lett. \textbf{98}, 067204 (2007).
\bibitem{flindt} T. Novotn\'y, A. Donarini, C. Flindt, and A.-P. Jauho, Phys. Rev. Lett. \textbf{92}, 248302 (2004). 
\bibitem{nov} T. Novotn\'y, A. Donarini, and A.-P. Jauho, Phys. Rev.
Lett. \textbf{90}, 256801 (2003).
\bibitem{sch2} M. G. Schultz, Phys. Rev. B {\bf 82}, 155408 (2010).
\bibitem{koenig96} J. Konig, J. Schmid, H. Schoeller, and G. Schon, Phys. Rev. B {\bf 54}, 
16820 (1996).
\bibitem{timm} C. Timm, Phys. Rev. B \textbf{77}, 195416 (2008).
\bibitem{gurv} S.A. Gurvitz and Y.S. Prager, Phys. Rev. B \textbf{53}, 15932 (1996).
\bibitem{abra} M. Abramovitz, I. Stegun, \textit{Handbook of Mathematical Functions with Formulas, Graphs, and Mathematical Tables}, Dover, 1964.
\bibitem{scholler} H. Schoeller and J. K\"onig, Phys. Rev. Lett. \textbf{84}, 3686 (2000).
\bibitem{maser} G. Rempe, F. Schmidt-Kaler, and H. Walther, Phys. Rev. Lett. \textbf{64}, 2783 (1990).
\bibitem{santa} D. H. Santamore, A. C. Doherty, and M. C. Cross, Phys. Rev. B {\bf 70}, 144301 (2004).
\bibitem{buks} E. Buks, E. Segev, S. Zaitsev, B. Abdo, and M. P. Blencowe, Eur. Phys. Lett. {\bf 81}, 10001 (2008).
\bibitem{woolley} M. J. Woolley, A. C. Doherty, and G. J. Milburn, Phys. Rev. B {\bf 82}, 094511 (2010).
\bibitem{o_connel} A. D. O'Connell, M. Hofheinz, M. Ansmann, R. C. Bialczak, M. Lenander, Erik Lucero, M. Neeley, D. Sank, H. Wang, M. Weides, J. Wenner, John M. Martinis, and A. N. Cleland, Nature \textbf{464}, 697 (2010). 
\end{thebibliography}
\end{document}